# Free electron screening mechanism of the shallow impurity breakdown in n-GaAs: evidences from the photoelectric Zeeman and cyclotron resonance spectroscopies

O. Alekperov[1] and E. Nakhmedov[1,2]

[1]Institute of Physics, Azerbaijan National Academy of Sciences,
H. Cavid ave. 33, AZ1143 Baku, Azerbaijan,

[2] Faculty of Physics, Moscow State University, Baku branch,
str. Universitetskaya 1, AZ-1144 Baku, Azerbaijan

(Dated: September 26, 2017)

## Abstract

A novel breakdown (BD) mechanism of shallow impurity (SI) under the electric field at low temperatures is suggested for $n - GaAs$ samples with the donor concentrations $N_D = 10^{14} \div 10^{16} cm^{-3}$ and the compensation degree $K = \frac{N_A}{N_D} = 0.3 \div 0.8$ with acceptors of concentration $N_A$ in the external magnetic fields up to $H = 6.5\ T$, oriented toward parallel or perpendicular to the external electric field. Diagnosis of the BD mechanism was performed by SI Zeeman (mainly from the ground state $1s$ to the $2p_{+1}$ and other excitation states) and cyclotron resonance photoelectric spectroscopy (PES) methods in the wide interval of the electric field including the BD region too. The obtained results reveal that the BD electric field $\mathcal{E}_{BD}$ does not correlate with $K$ and the carriers mobility $\mu$ of the samples, which contradict to the well-known impact ionization mechanism (IIM). A serious discrepancy with IIM is that, $\mathcal{E}_{BD}$ does not almost depend on the magnetic field up to $H = 6.5\ T$ when $\mathcal{E} \| H$, though the SI ionization energy increases two times. The cyclotron resonance (CR) measurements show that the line width does not depend on the electric field for $\mathcal{E} < \mathcal{E}_{BD}$ indicating the lack of free carriers (FC) heating in contradiction with IIM. A considerable decrease of the free carriers' capture cross section (CCS) by ionized SI centers with a subsequent increase in the FC concentration $n$ is observed by means of PES investigation of the $1s \rightarrow 2p_{+1}$ and CR lines in the electric fields $\mathcal{E} \leq \mathcal{E}_{BD}$ and at different magnetic fields, applied along $H \| \mathcal{E}$ or perpendicular $H \perp \mathcal{E}$ to the electric field. The slope of the $1s \rightarrow 2p_{+1}$ line intensity on the electric field for $\mathcal{E} \| H$ does not depend on magnetic field, which is valid for $\mathcal{E}_{BD}$ too. Various effects determined in the PES measurements at $\mathcal{E} = \mathcal{E}_{BD}$, such as a drastic narrowing the $1s \rightarrow 2p_{+1}$ and CR lines, a shift of the CR line to higher magnetic fields and disappearing of the lines to higher excited SI states, were clarified to be a result of screening of SI Coulomb potential by free carriers. The FC screening at the BD reduces the potential fluctuation and its influence to the PES line-shape of $1s \rightarrow 2p_{+1}$ and other excited states. It is shown that an increase in the FC concentration reduces the CCS, which can be assumed as the main factor along with the increase in the ionization coefficient for the SI breakdown in the electric field. The screening length $r_s$ of the SI Coulomb potential decreases with increasing the FC concentration, reducing the CCS; the latter seems to vanish completely at $r_s = a_B^*$ ($a_B^*$ is the effective Bohr radius), when high screening results in vanishing of all the bound states of the Coulomb potential. Note that this limit is similar to the Mott transition. Many experimental facts and our calculation of the CCS support the suggested mechanism for the SI breakdown. The well-known IIM is valid for samples with SI concentrations $N \ll 10^{13} cm^{-3}$, and takes place at very high electric fields.

## I. INTRODUCTION

The gallium arsenide is one of the most utilized semiconductor in the modern electronics technology. $n - GaAs$ epitaxial layers are the widely used hetero-junctions for the investigation of the comprehensive class of 2D modern electronic and spintronic phenomena and for fabrication of different devices on their basis such as Gann diodes, photodetectors operating in the wide range of frequency, high frequency field transistors etc. In order to improve the electro-physical characteristics of these devices, fabricated by using high-purity semiconductors, it is necessary to know the chemical nature and the relative concentration of residual impurities. Many of these devices operate at low temperatures and sufficiently high electric fields when shallow impurities (SI) breakdown (BD) takes place (see, for review, e.g. [1, 2]). Therefore more careful investigation of SIBD mechanism in $n - GaAs$ is essential.

The low temperature sub-millimeter wave photoelectric spectroscopy (PES) of SI is the most sensitive method for identification of the SI contents in semiconductors [3]. A significant information on SI can be obtained from the photoconductivity spectra line-width and the line broadening mechanism. The line-width of the SI PES was shown [4, 5] to be determined by the concentrations of the major and compensated impurities as well as their distribution. The difference of ground state energies $E(1s)$ for different SI atoms, which is called a chemical shift (CS) or central cell correction, is a result of deviations ($\Delta V$) of SI Coulomb potential $V(r) = e/(\epsilon_0 r) + \Delta V$ at small distances from pure Coulomb form. The value of the CS for SI in $n - GaAs$ can be estimated to be $\Delta E \sim (a_0/a_B^*)^2 \epsilon_0 Ry^* \approx 0.1 meV$, where $a_0$ and $a_B^*$ are the unit cell size and the effective Bohr radius, correspondingly, and $Ry^*$ is the effective Rydberg. This means that the energies of all SI in $n - GaAs$ lay in the interval of $Ry^* \pm \Delta E = (5.7 \pm 0.1) \ meV$ therefore CS correction to higher excited states of SI ($E(2p_{+1})$ and $E(3p_{+1}),...$) is much smaller than that for $1s$ state and can be neglected. For the purest semiconductors with donor concentrations of $N_D < 10^{14} \ cm^{-3}$, the photoexcitation lines of impurities, the intensity of which is proportional to the concentration of the corresponding impurity, are broadened due to the different values of the CS, and the broadening value for $n - GaAs$ samples is the same order as the energy distance between impurities. Each impurity atom is considered in this case to be isolated, the line width broadening is determined only by the charged impurity mechanism. As we will show in the following Section, the quadratic Stark effect and the quadrupole-gradient shifting of neutral impurities' levels are responsible for the inhomogeneous broadening of the charged impurities. Note, that such a small value of CS, which is the same order as the impurity photoexcitation line width is inherent to the most of $A_3B_5$ semiconductors ($InSb$, $GaAs$, $InP$). However the difference of SI ionization energies for different impurities in $Ge$ and $Si$ are in the order of $Ry^*$. That is why the CS in these materials does not affect to the line width.

For the $n - GaAs$ samples with the SI concentrations higher than $10^{14} \div 10^{15} cm^{-3}$ the picture is significantly different [6]. First of all, our experiments have determined that the line width for these samples does not correlate with the SI concentration, so that for a sample with smaller concentration the line width may be larger than that of more doped samples (see, Table I). Our investigations show that the line width of these samples is determined not only with the SI concentration but with the potential fluctuation due to the inhomogeneous SI donors and acceptors distribution too. Such an inhomogeneous SI distribution creates a potential fluctuation, which gives an additional contribution to the line width. At lower temperatures, close to zero, the electrons are captured by impurities, providing a correlated distribution of electrons, which is realized by minimizing the total Coulomb energy of the system of the charged donors, acceptors and electrons [7]. At higher temperatures, when $T \gg T_c = \frac{e^2}{\epsilon_0 r_m k_B}$ with $r_m = (4\pi/3N_I)^{1/3}$ being the mean distance between charged impurities, the electrons are activated, and their distribution becomes random. The $1s \to 2p_{+1}$ transition line width in this case depends on the charged impurity concentration $N_I = 2N_A$, and does not depend on the compensation $K = N_A/N_D$. Instead, the line width in the real experiments, which are realized at lower temperatures, the electrons distribution is correlated, since the acceptors are charged, taking an electron from nearest donor [5], and the former strongly depends on $K$. A transformation from the correlated to the random distribution can be reached not only by increasing temperature but also applying an external electric field.

The impurities in a strongly inhomogeneous sample gather into cluster with higher impurity concentration. Hall measurement yields only mean value of the impurity concentration. Since these clusters are optically active, they are responsible for the radiation absorption. The impurity orbitals of the neighboring donors in the cluster are overlapped. Therefore, the $1s \to 2p_{+1}$ transition is allowed not only to the excited $2p_{+1}$ state of the owner atom but to neighboring impurity $2p_{+1}$ states too. Note that just such kinds of transitions cause potential fluctuations broadening of $1s \to 2p_{+1}$ transition line-width. The line-width for these transitions depends on the radiation intensity also [8, 9]. The transition between different impurity states may provide an additional inhomogeneous broadening mechanism for inhomogeneous impurity distribution at higher concentrations. Therefore, identification of the lines becomes difficult due to inhomogeneous line broadening, which increases significantly at the SI concentrations higher than $10^{14} \div 10^{15} cm^{-3}$.

It is clear that the width of the impurity PEL has to be reduced in order to increase the resolution of the PES, which needs an investigation of the inhomogeneous broadening mechanisms of PES of SI. An additional inter-band illumination of samples causes the some narrowing of SI PES line width [10] due to decreasing of charged impurities concentration. Nevertheless, our experiments showed that a narrowing of the $1s \rightarrow 2p_{+1}$ line in PES by inter-band illumination is effective only in samples with high compensations $K = N_A/N_D > 0.5$. We found that the SI PES line widths are considerably narrowed at the SI breakdown electric field.

The aim of this work is to investigate a mechanism of the electric field breakdown and of the broadening of the photoelectric excitation spectra lines of shallow impurities and the CR in $n - GaAs$ samples with the donor concentrations $N_D \geq 10^{14} cm^3$ and intermediate compensations. Effects of the electric breakdown to the line shapes of $1s \rightarrow 2p_{+1}$ transition allow one to develop a method of determination of the impurity contents and their relative concentration in the highly doped samples.

According to the existence theories [1, 2, 11–14], the SI breakdown occurs by means of the impact ionization of the neutral impurities by hot electrons, heated in the electrical field to energies more than $2E_i$. The idiom of 'impact ionization' was introduced in physics firstly by Townsend [15] in order to explain the effects of an electrical discharging in gas. Further, Ioffe was argued [16] that the impact ionization may be a reason of the electric breakdown in solid insulators. The electron multiplication effect in $Ge$ and $Ga$ $p-n$ junctions [17] as well as in $n-Ge$ and $p-Ge$ [18, 19] was attributed to the impact ionization of impurities by free charge carriers. We present in this work a new mechanism of the SI PES line broadening and of the electric field breakdown mechanism of SI in a $n-GaAs$ sample, consisting of the charged impurities screening with free electrons, which is an alternative one to the impact ionization. This SI BD mechanism takes place at higher concentrations of the SI when $N_D > 10^{13} cm^3$ A brief description of this mechanism was mentioned firstly in Ref. [20]. The non-homogeneity degree of the impurity distribution in samples with similar concentrations of the major and the compensated impurities was shown to be determined by the width of $1s \to 2p_{+1}$ SI PES in the linear part of the current-voltage characteristics (CVC) and by the electric field dependence of the width in the pre-breakdown electric field.

The main scattering mechanism of electrons in the samples under our experimental condition is scattering on the ionized impurities. However, the investigation of the CR line width at the pre-breakdown electric fields shows no correlation with ionized impurity concentration as it must be $\omega_c \sim H\mu$. A reason of the CR line width broadening and different values of the cyclotron mass under the same experimental conditions is established to be a potential fluctuation due to non-homogeneous distribution of the impurities. Drastic increase of the free electron concentration at BD results in two kind of CR line shift; the first shift is connected with influence of the plasma oscillation as a result of enhancement of the free electron, and it increases with decreasing the magnetic field. The second shift, which takes place at higher magnetic fields, increases with magnetic field due to decrease of influence of fluctuation potential on CR line-shape as a result of free carriers screening. Note that $\Delta H < 0$ at the plasma shift, while $\Delta H > 0$ at the screening shift of the CR line. It is worthy to note that a comparison of the CR line-shapes between breakdown and pre-breakdown electric fields allows us to characterize a degree of the non-homogeneous distribution of the impurities. The dependence of the CR line width on the electrical field around the breakdown $\mathcal{E} \approx \mathcal{E}_{BD}$ is experimentally observed to be non-homogeneous: the CR line is considerably narrowed at the beginning part of the breakdown, whereas the line width increases and exceeds several times its own pre-breakdown value. The analysis of the experimental data establishes that the breakdown takes place by means of the current filament formation in the sample. At the beginning stage of the breakdown, the released free electrons in the current filament screen the charged impurities and reduce their influence on the cyclotron motion of the electrons. The further increase in the electron concentration with the current, the electron-electron correlations dominate over the electron-impurity interactions, which lead to the broadening of the CR line width. The reason of the increase in the CR intensity with the electric field is established to be an increase in the free electron concentration in the first Landau level and in the life-time of the photoexcited electrons in the first Landau level due to reduction of the capture coefficient at the impurity centers.

The novelties can be summarized as follows:

1. The PES of the Zeeman lines of $n - CaAs$ samples with SI donors concentrations

smaller than that corresponding to Mott transition $N_D \approx 10^{14} \div 10^{16} \ cm^{-3}$ and with the intermediate compensations $K = N_A/N_D = 0.3 \div 0.8$ at the breakdown electric field undergoes to drastic narrowing of the PES lines $1s \rightarrow 2p_{+1}$ and $1s \rightarrow 3p_{+1}$;

2. All PES lines, corresponding to transitions from the ground state of the donors to the quasi-discrete excited states higher than $3p_{+1}$, disappear at pre-breakdown region of CVC;

3. Drastic narrowing of the CR line-width in PES occurs at break-down region of CVC of samples;

4. All these experimental facts can be explained by screening of the charged impurity potential by free electrons released in the breakdown process. Calculations of the $1s \rightarrow 2p_{+1}$ transition line-shape for the two main line broadening mechanisms, namely the quadratic Stark and the quadrupole-gradient effects, by replacing the surrounding the neutral impurity charged impurities Coulomb potential with the screened Coulomb potential, $(e/\epsilon_0 r) \exp(-r/r_s)$, explain the narrowing of the FES line. The calculations yield that the $1s \rightarrow 2p_{+1}$ line is narrowed several times for the value of the screening length $r_s = (4\pi e^2 n^{-1} \epsilon_0^{-1})^{1/2}$ corresponding to free electron concentration $n$ smaller than the neutral impurity concentration;

5. Two different kinds of shifts of the CR line at the break-down takes place, both of them are connected with abrupt increase of free carrier concentration $n$ at breakdown. The first shift ($\Delta H_{CR} < 0$), which increases with decreasing of CR magnetic field value $H_{CR}$ is due to free carriers plasma influence on CR line-shape. The second shift ($\Delta H_{CR} > 0$), which increases with $H_{CR}$ is due to the screening of charged impurities potential fluctuations with free carriers.

6. Two kinds of the CR are observed at small electric fields corresponding to the linear region of CVC in PES. The first is the free carriers CR line, at slightly smaller magnetic field with very low intensity in comparison with that of broader line of the pseudo-free carriers CR of electrons localized on fluctuation of potential CR (FPCR). With increasing the electric field, activation of carriers from the potential fluctuation to the zeroth Landau level causes strong increase in the free carrier CR in comparison to FPCR broad line. Such kind of CR electric field dependence is typical for samples with inhomogeneous distribution of impurities. So this fact can be used in characterization of impurity distribution in samples of $n - GaAs$ with nearly equal $N_D$ and $K$.

7. The intensity $\Delta\sigma/\sigma$ of photoelectric excitation spectra $1s \rightarrow 2p_{+1}$ and CR lines was shown to increase with the pre-breakdown electric field due to the decrease in the capture cross-section $\alpha(\mathcal{E})$ and increase in the coefficient of thermal ionization from excited states $\beta(\mathcal{E})$. So the concentration of free electrons was shown increases exponentially at pre-breakdown electric fields.

Our experiments have been done at different values of the magnetic field, $H = 37 \; kOe$, and $H = 61 \; kOe$, and for two mutual orientations of the electric and magnetic fields, $\mathcal{E}\|\mathbf{H}$ and $\mathcal{E} \perp \mathbf{H}$. The experimental results confirm that a reason of an increase in the $1s \rightarrow 2p_{+1}$ line intensity in the electric field is a decrease in the capture cross-section of the excited electrons by the impurity centers. Furthermore, the capture cross-section for given values of the electric and magnetic fields depends on their mutual orientations: it takes a minimal value for $\mathcal{E}\|\mathbf{H}$ and maximal value for $\mathcal{E} \perp \mathbf{H}$.

## II. EXPERIMENT AND MEASUREMENT METHODS

The experiments were provided in six samples of $n - GaAs$ epitaxial layers with the residual donor concentrations ($10^{14} \; cm^{-3} < N_D < 10^{16} \; cm^{-3}$) and compensation degree $K = 0.3 \div 0.8$, grown on the high-resistive $p - GaAs$ substrate by using the liquid-phase epitaxy (LPE) and vapor-phase epitaxy (VPE) methods. The samples of thickness $15 \div 150 \; \mu m$ have the surface area of several $mm^2$. Hall measurements of the samples [21] are provided to determine $N_D$, $K$ and the mobility $\mu$. The sub-millimeter laser magnetic spectrometer was used to investigate the SIPES and CR line-shapes. Sub-millimeter gas-discharge lasers operating on flowing regime of $H_2O + H_2$, $D_2O + D_2$ and $HCN(CH_4 + N_2)$ vapors were used as a radiation source with the radiation wavelengths of $118.6 \; \mu m$ , $78.4 \; \mu m$, $84.3 \; \mu m$, and $337 \; \mu m$. Photoconductivity spectra of CR and SIPES were registered at fixed quantum energies of lasers as a function of magnetic field with scanning of magnetic field intensity up to $H = 65 \; kOe$ of superconducting solenoid, setting in liquid helium cryostat at $T = 4.2 \; K$. Cross-modulation method was used in registration of photoconductivity with modulation of radiation intensity at frequency $750 \; Hz$. Alternating signal of photoconductivity was measured from load resistor $R_L$ series-connected with sample with resistivity of $R_S$ by using of the lock-on method. Voltage drops from $R_L$ and $R_S$ were used in $CVC$ registration respectively. Line-shapes of CR and SIPES were studied at different regions of samples' CVC. At electric fields smaller than the breakdown one $\mathcal{E} < \mathcal{E}_{BD}$, the measurements were done in constant voltage regime, when $R_L \ll R_S$ for different values of the electric field. In this case the photosignal is proportional to $\Delta\sigma$ under the radiation. At pre-breakdown and at "candle"-like region $\mathcal{E} \approx \mathcal{E}_{BD}$, the measurements are performed in constant current regime, when $R_L$ is much greater than sample resistivity $R_S$, $R_L \gg R_S$. In constant voltage regime PC signal is proportional to the change of conductivity $\Delta U \sim \Delta\sigma$ under the radiation and as a result PC line shape corresponds to real line-shape at $\mathcal{E} < \mathcal{E}_{BD}$. However, it is not suitable at $\mathcal{E} \approx \mathcal{E}_{BD}$ due to the instability of electric parameters of sample. Constant current condition gives the possibility to detect line shapes at candle-like region of CVC at different free carrier concentration. In this case PC of sample $\Delta U \sim \Delta\rho = \frac{1}{\sigma_{dark}} - \frac{1}{\sigma_{light}} = (e\Delta n\mu)/(\sigma_{dark}^2 + \sigma_{dark}e\Delta n\mu)$, where $\sigma_{dark}$ is a dark conductivity. At small dark conductivity $\sigma_{dark} \ll e\Delta n\mu$ (which correspond to small electric field) PC signal would have a saturation with increasing of $\Delta n$, as a result PC spectra do not correspond to true line-shape. In opposite case, which corresponds to the breakdown region ($\sigma_{dark} > e\Delta n\mu$), photo-signal is proportional to the number of resonance transition $\Delta n$, and it corresponds to the true line-shape. The analysis of PC signals $\Delta U$ at breakdown region shows that the saturation of $\Delta U$ on the radiation intensity takes place at sufficiently high $\Delta n$ comparable with impurity concentration. This means that the line widths of CR and $1s \rightarrow 2p_{+1}$ transitions obtained at the breakdown in the current regime are only broader than the true line shapes. This is important for us because our conclusions partially are based on strong decreasing of these line-shapes at SI breakdown. It is obvious that at constant voltage only and constant current conditions it is possibility to observe N-and S-like regions of CVC of samples respectively. Nevertheless, CVC of samples obtained at constant current regime should not show a $S$-like region as illustrated in CVC at different magnetic fields in Fig.1. Note that in the case of CVC recording in constant voltage regime at $\mathcal{E} \approx \mathcal{E}_{BD}$, the voltage drop switches from sample to load resistor at the beginning of the breakdown, and this causes decreasing of $\mathcal{E}$ on sample; and as a result of this, an external voltage drop returns back to sample after time $\tau_{cap}$, when the excited free carriers are captured back to impurity sites again. This corresponds to the first cycle of the oscillation, which is known as low frequency current oscillations (LFCO) on sample at impurity breakdown and other switching effects. The amplitude of these oscillations decreases with increasing the load resistance, and it disappears at strong constant current regime. The points on CVC line, obtained at strong constant current condition correspond to load resistance value nearly 10 times greater than the resistivity of sample at the beginning of the breakdown. As it is seen from Fig. 1, fluctuations of voltage in the sample decrease at strong current regime.

| samples | $N_D$ $(\times 10^{14} cm^{-3})$ | $K = \frac{N_A}{N_D}$ | $\mu$ $(\times 10^4 \frac{cm^2}{V \cdot s})$ | $\Delta E(meV)$ $H \approx 36 kOe$ | $\Delta E(meV)$ $H \approx 61 kOe$ | $\delta(meV)$ | $\delta/\Delta E$ |
|---|---|---|---|---|---|---|---|
| 1 VPE | 4.33 | 0.52 | 10 | 0.064 | 0.07 | 0.214 | 3.06 |
| 2VPE | 7.35 | 0.65 | 7.3 | 0.07 | 0.08 | 0.270 | 3.38 |
| 3LPE | 8.67 | 0.53 | 7.3 | 0.15 | 0.16 | 0.271 | 1.69 |
| 4VPE | 1.46 | 0.38 | 5.7 | 0.085 | 0.09 | 0.297 | 3.03 |
| 5LPE | 4.22 | 0.36 | 8.6 | 0.136 | 0.14 | 0.194 | 1.39 |
| 6VPE | 4.31 | 0.52 | 10 | 0.062 | 0.065 | 0.214 | 3.29 |
| 7VPE | high purity | – | – | 0.012 | 0.008 | | |

TABLE I: Characteristics of different $n-GaAs$ samples, grown by liquid-phase epitaxy (LPE) and vapor-phase epitaxy (VPE); the experimentally measured $\Delta E$ and theoretically predicted $\delta$ values of the line widths of $1s \rightarrow 2p_{+1}$ donor lines at different cyclotron frequencies.

The parameters of the samples and their line width characteristics at different magnetic fields are given in Table I. The line width $\Delta E$ was measured at half-height of the line in magnetic field units $\Delta H$ and then it was transformed into energetic units $\Delta E = \left( \partial E_{1s \rightarrow 2p_{+1}}/\partial H \right) \Delta H$. The rate of $1s \rightarrow 2p_{+1}$ transition energy increase in magnetic field $\partial E_{1s \rightarrow 2p_{+1}}/\partial H \approx 0.18 \ meV/kOe$ was determined at two different values of the magnetic field $H \approx 36.5 \ kOe$ and $H \approx 60.5 \ kOe$. The preliminary CVC measurements have been done at the magnetic fields $\mathbf{H}$ corresponding to the top of the photoexcited spectra as well as in the electric fields $\mathcal{E}$ crossed and parallel to the magnetic field. CVC of a typical sample, measured in the crossed magnetic and electric fields, for different magnetic fields is presented in Fig. 1 at either fixed voltage (solid lines) or fixed current (dotted lines) regimes. A jump in the CVC at the electric field $\mathcal{E} = \mathcal{E}_{BD}$ indicates a breakdown the neutral impurities. It is worthy that $\mathcal{E}_{BD}$ increases linearly with magnetic field at $\mathcal{E} \perp \mathbf{H}$, whereas it does not depend practically on $H$ at $\mathcal{E} \| \mathbf{H}$, as it is shown in Fig. 2.

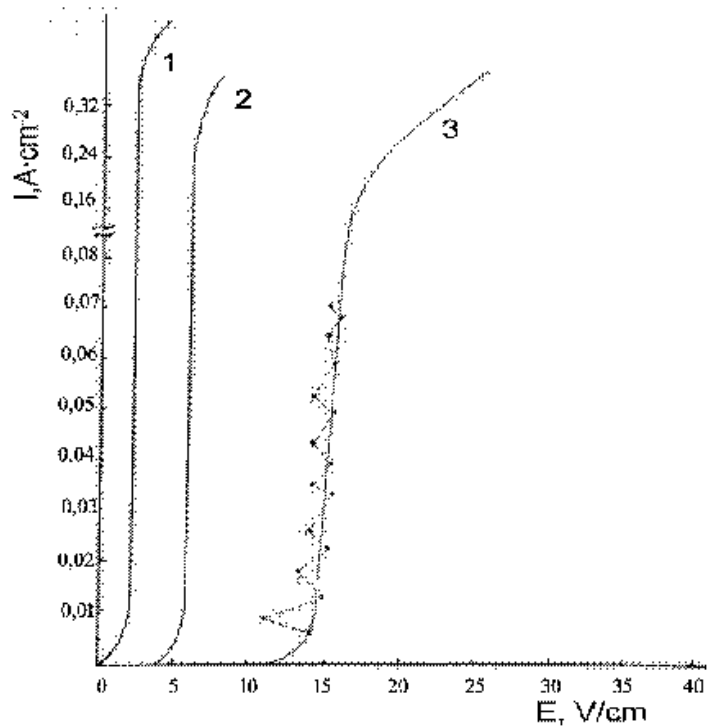

FIG. 1: CVC of $n - GaAs$ samples measured at different magnetic fields in the strictly constant current regime at $R_L > 100\ R_S$ (solid curve) and in the pseudo-current regime when $R_L \approx 10 R_S$ (dotted curve).

## III. INFLUENCE OF THE ELECTRIC FIELD BREAKDOWN ON THE SI PHOTOELECTRIC SPECTROSCOPY

According to the conventional explanation [3–5, 7], the main mechanism of the SI photoexcitation line broadening in magnetic field is the quadratic Stark effect and the electric field gradient of the quadrupole interactions of charged impurities. Our experiments show that there exists an additional line broadening mechanism in the samples with high impurity concentrations, which is caused by the potential fluctuations. As it is seen from Table I, there is no correlation of the PES line width neither with SI concentration nor with the degree of the compensation for $N_D \approx 10^{14}$ to $10^{15}\ cm^{-3}$ and $K \approx 0.3$ to $0.8$. This evidence can be understood, providing that an inhomogeneous SI distribution plays an essential role in the line broadening. At high impurity concentrations, the impurity states are broadened forming an impurity band. In this case, a transition takes place not only from the $1s$ ground state of a given neutral donor to the $2p_{+1}$ excited state to the same atom but to the nearest-neighboring charged donors too. Since high SI concentration makes it possible a partially wave function overlapping of the excited states. The distribution function of the charged SI, randomly distributed in the sample, was proposed to be [22] in the Gaussian form

$$P(e\varphi) = \frac{1}{\delta\sqrt{2\pi}} exp\left(-\frac{e^2\varphi^2}{2\delta^2}\right),\qquad(1)$$

where $\delta$ is half-width of Gaussian distribution, and $\delta = 0.29e^2 N_D^{1/3} K^{1/4} \epsilon_0^{-1}$ characterizes the theoretically predicted value of the line width $E_0$; $\delta$ increases with the impurity concentration and the compensation degree. This line broadening mechanism dominates at $N_D > 10^{14}\ cm^{-3}$ in $n - GaAs$, [6]. The values of $\delta$ for the samples under investigation with $\epsilon_0 = 12.5$ are given in Table I. Although the magnitudes of $\delta$ are of the same order as $\Delta E$, they are 1.5 and 3 times greater than the experimental values of the $1s \to 2p_{+1}$ line widths for VPE and LPE samples, correspondingly.

A discrepancy between the theoretical and experimental values of the PES line width seems to be explained by screening of charged impurity potential by electron gas. The screening of $\delta$ can be qualitatively described as

$$\delta^* = \delta \exp\left(-\frac{r_m}{r_s}\right),\qquad(2)$$

where $r_m = \left(\frac{4\pi}{3N_I}\right)^{1/3}$ is the mean distance between the charged impurities $N_I \approx 2N_A$ and $r_s = (\epsilon_0 k_B T/4\pi e^2 n)^{1/2}$ is Debye screening radius. For $N_A \approx 4 \times 10^{14}\ cm^{-3}$ one can estimated $r_m \approx 13.5\ a_B$, and $r_s \approx 10\ a_B$ at Helium temperature, provided that $\sim 10\%$ of the acceptors is ionized. These values reduce the value of $\delta$ to $\delta^* \approx 0.27\delta$, which is in consistence with the experimentally measured value.

The requirement that $\delta$ and the experimentally measured half-line width $\Delta E$ take the same values imposes a condition that the intra-impurity transitions have to be neglected at all, and inter- impurity transitions not only between the nearest-neighboring donors but between any two impurities have to be realized with equal probability. $\delta$ takes larger values yielding wider line width where a local donor concentration is higher in the inhomogeneous impurity distribution even for a sample with smaller $N_D$. Therefore, the value $\Delta E/\delta$ as well as $(\Delta E - \delta)$ can serve as a measure of SI inhomogeneous distribution degree in samples with nearly equal values of $N_D$ and $K$. One concludes that if a sample with higher SI concentration has smaller line width then it has more homogeneous distribution of the donors. The value of $\delta/\Delta E$ is a suitable parameter for comparison of a sample quality, so that the higher $\delta/\Delta E$ the better is the quality of a sample in comparison with other ones with less homogeneous SI distribution. This parameter is shown in Table I for $H \approx 61\ kOe$ only, since neither $\delta$ nor $\Delta E$ depends on magnetic field.

The energy distance or gap between the $1s$ ground state and $2p_{+1}$ excited state of electron in the impurity increases with magnetic field, and the resonance $1s \rightarrow 2p_{+1}$ transition occurs when the gap becomes equal to the radiation energy. At $H > 20 kOe$, which corresponds to the intermediate magnetic fields $\gamma = \frac{\hbar\omega_c}{2Ry^*} = 0 \div 1$ with $\omega_c = eH/m^*c$ being the cyclotron frequency ($\gamma = 1$ in $n - GaAs$ corresponds to $H \approx 65\ kOe$), the $2p_{+1}$ level of shallow impurities in $n - GaAs$ lies higher than the zeroth Landau level ensuring a transition for the photoexcited electrons to the conduction band (see, Fig. 3). Therefore, there is not a necessity to argue other mechanism, like field induced tunneling or impact ionization from excited $2p_{+1}$ state to explain a generation of the PES signal.

It is worthy to note that the line shape of high purity samples in the PES is asymmetric (Fig. 3 in [6]). The line width at lower energy or higher magnetic field side of the $1s \rightarrow 2p_{+1}$ transition is considerably wider in accordance with the quadratic Stark effect broadening mechanism. Instead, all other samples have symmetric line shape. The line width for the $1s \rightarrow 2p_{+1}$ transition of the purest sample is 1.5 times narrowed when the magnetic field increases from $H \approx 36\ kOe$ to $\sim 61\ kOe$, which confirms also a validity of the quadratic Stark broadening mechanism [4].

PES of the SI donors at the breakdown electric field is measured at the constant current regime, which ensures stability of the electrical parameters of samples during PES registration. The impurity PES line at sufficiently higher current density, corresponding to the deep breakdown regime at $\mathcal{E} \approx \mathcal{E}_{BD}$ have the following features:

- The transition lines $1s \rightarrow 2p_{+1}$ and $1s \rightarrow 3p_{+1}$ are narrowed from 5 up to 10 times with increasing the current density for all samples under investigation (see, Fig. 4);

- The $1s \rightarrow 2p_{+1}$ line acquires an asymmetric form while a narrowing takes place, and the half-width becomes wider at higher magnetic fields. These behaviors are similar

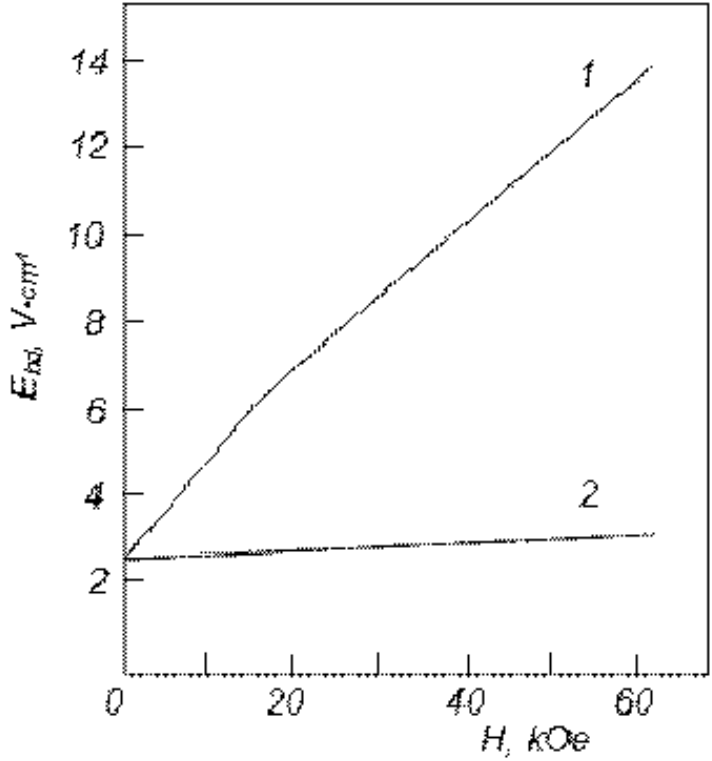

FIG. 2: Dependence of the breakdown electric field $\mathcal{E}_{BD}$ on the magnetic field $H$ for different $n - GaAs$ samples when the electric $\mathcal{E}_{\mathbf{BD}}$ and magnetic $\mathbf{H}$ fields are (1) perpendicular $\mathcal{E}_{\mathbf{BD}} \perp \mathbf{H}$ and (2) parallel $\mathcal{E}_{\mathbf{BD}} \| \mathbf{H}$ each other. A weak dependence of $\mathcal{E}_{BD}$ on $H$ may be caused by an error in the setting of a rotation angle of the sample.

to those for a pure sample;

- the lines in the photoelectric spectra, corresponding to the transitions from the $1s$ ground state to the quasi-discrete Boyle-Howard's $\{N, M, \nu\}$ states [23], determined by the main quantum number $N$ corresponding to the Landau levels, the magnetic quantum number $M$, and the number $\nu$ of the bounded discrete Coulomb levels (the Coulomb quantum number), disappear under $N \geq 1$ with increasing the breakdown current (Fig. 5), i. e. apart from the $1s \to 2p_{+1}$ the transition line $1s \to 3p_{+1}$ disappears too;

- Intensity of the spectra at the tails strongly decreases;

- The new lines in the PES emerge due to narrowing of the $1s \to 2p_{+1}$ line, which

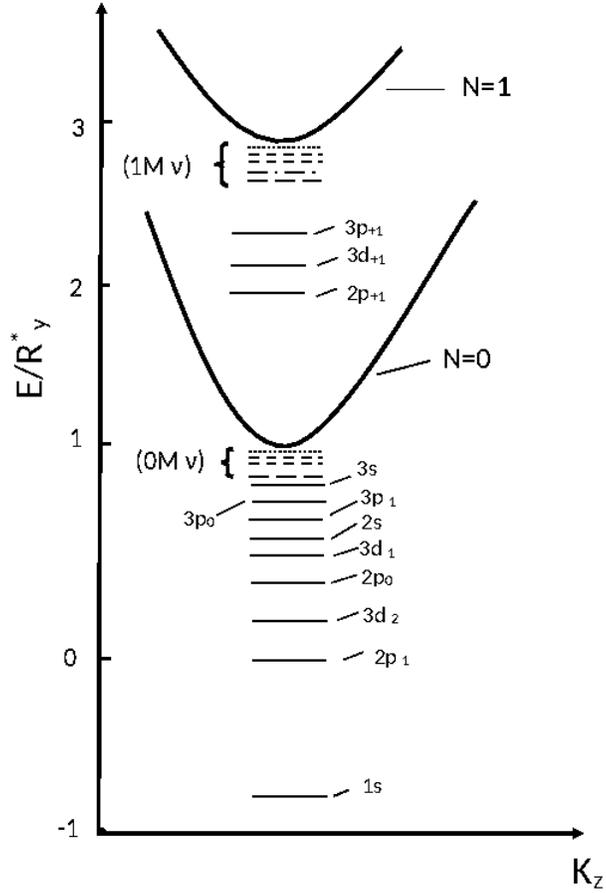

FIG. 3: The shallow donor levels in $n - GaAs$ in the presence of the magnetic field corresponding to $\gamma = 1$. The $2p_{+1}$ excited impurity state lies higher than the $N = 0$ zeroth Landau level under $N = 1$. The zero energy is labeled from the bottom of the conduction band at $H = 0$.

correspond to residue SI donors with smaller concentrations, (see, Fig. 4);

- narrowing of the $1s \rightarrow 2p_{+1}$ line results in an appearance a fine structure of each donor line in consequence of interactions of an electron spin with the external magnetic field (see, Fig. 4).The revealing of the spin-splitting in the PES line makes the line structure very similar to those observed in the purest samples.

The above specified features of the PES at the breakdown electric field can be explained by assuming that the free electrons, released at the breakdown, screen the Coulomb potential of charged impurities. Note that a screening influences to excited states of neutral donors much more, resulting in their disappearance.

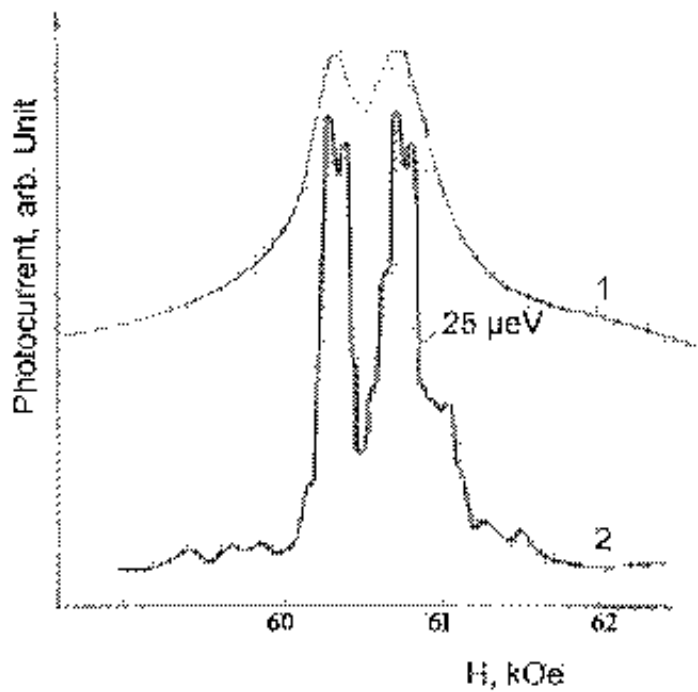

FIG. 4: Shape of the $1s \rightarrow 2p_{+1}$ line of $n-GaAs$ sample 6 at the radiation energy of $E = 14.7\ meV$ and $T = 4.2\ K$, $K = 0.5$ for (1) $\mathcal{E} < \mathcal{E}_{BD}$ in the constant field regime and (2) $\mathcal{E} \approx \mathcal{E}_{BD}$ in the constant current regime at $J = 17\mu A$. The PES lines are strongly narrowed at the breakdown, the fine structure of each line and new lines with smaller intensities appear at the tails of the main line.

The estimations show that the line broadening due to the excitation finite life-time is negligibly small, $\sim 10\ \mu eV$ [24], in comparison to the experimentally observed data even in a ultra-pure $n-GaAs$ samples. Therefore, the homogeneous broadening mechanism can be neglected for the samples under investigation.

The non-homogeneous PES broadening is determined as a shift between the transition energy of the impurity states in crystal and that of an isolated hydrogen-like impurity atom. In this case the PES line- shape is determined by statistical distribution of the optical transitions from the ground state to the excited ones for all impurity atoms. The non-homogeneous broadening of the PES line can be caused by [4, 5, 25] (i) the interactions between the neutral impurity atoms, (ii) quadrupole-gradient interactions of the charged and neutral impurity atoms, and (iii) the quadratic Stark effect of the electric field created by the randomly distributed charged impurities on neutral impurities. Note that the linear Stark effect is absent in magnetic field [4].

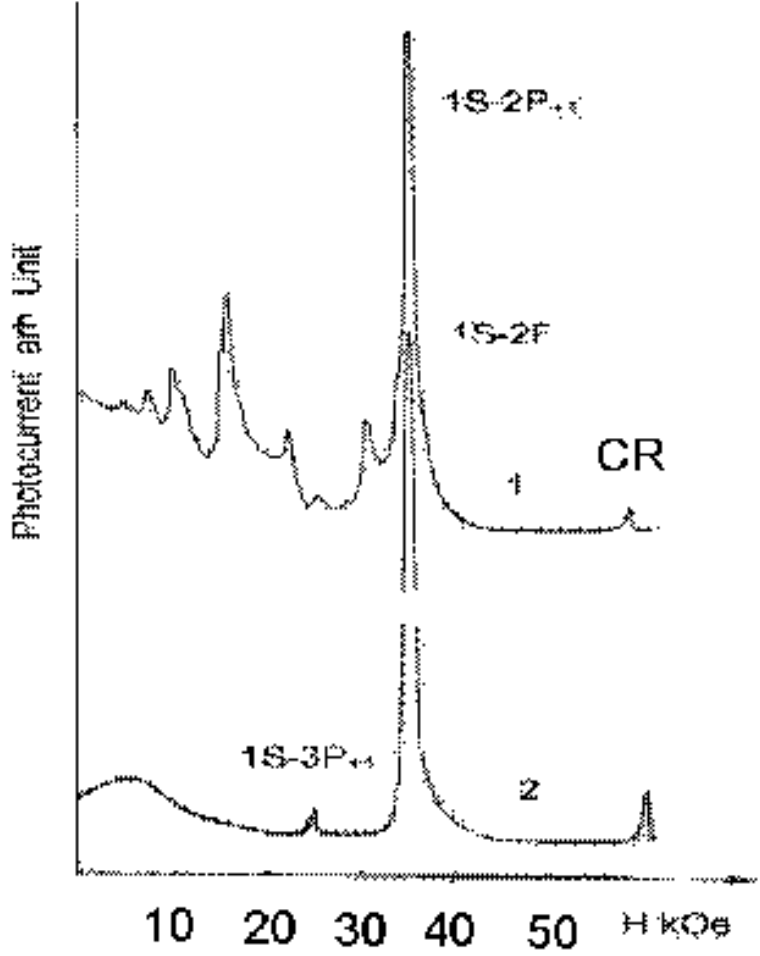

FIG. 5: Photoelectric spectra of $n-GaAs$ samples at $\mathcal{E}_{BD} = 10.45\ MeV$ and $T = 4.2\ K$ for (1) $\mathcal{E} < \mathcal{E}_{BD}$ and (2) $\mathcal{E} \approx \mathcal{E}_{BD}$.

Interactions between the neutral donors have a dipole-dipole interaction origin, which broaden the line width symmetrically in comparison with a unperturbed transition energy. The dipole-dipole interactions mediated half-width of the line in the absence of the magnetic field is given [3] by $\Delta_{d-d} \approx 6N_D^0 a_B^{*\,3}\epsilon_i$, which can be estimated to be $\Delta_{d-d} \approx 2 \div 20\mu eV$ for the neutral donor concentration $N_D^0 \sim (10^{14} \div 10^{15}) cm^{-3}$. This value in the intermediate magnetic field, $\gamma \approx 0.5 \div 1.0$, is much more smaller due to squeezing of the electronic wave functions in the excited states.

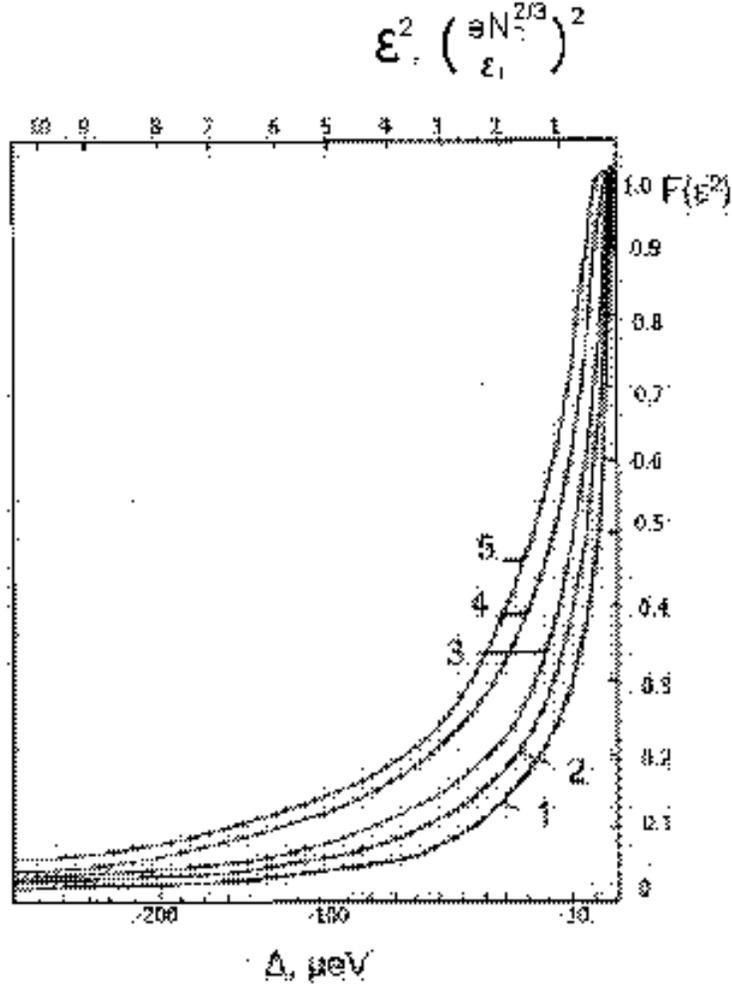

FIG. 6: Shape of the $1s \to 2p_{+1}$ line, determined by the quadratic Stark effect at $H = 0$, $N_D a_B^3 = 4 \times 10^{-4}$, $K = 0.5$, $T = 0$, for different impurity potentials: (1) $r_s = 10\ a_B^*$; (2) $r_s = 20\ a_B^*$; (3) $r_s = 30\ a_B^*$; (4) $r_s = 40\ a_B^*$; and (5) Coulomb potential. $\mathcal{E}^2$ is given in unit of $\left(\frac{eN_D^{2/3}}{\epsilon_0}\right)^2$.

We show that the distribution functions of quadratic electric field $P_1(\mathcal{E}^2)$ and of the electric field gradient $P_2\left(\frac{\partial \mathcal{E}_z}{\partial z}\right)$, which determine the quadratic Stark effect and the gradient broadening of PES line correspondingly, are squeezed due to the screening of the Coulomb potential of surrounding charged impurities by free electrons. Calculations of the distribution functions of $\mathcal{E}^2$ and $\frac{\partial \mathcal{E}_z}{\partial z}$ on the neutral donors are performed according to the method developed by Shklovskii and Efros [26]. In order to realize the ground state of a doped compensated semiconductor, the coordinates of $N$ donors and $KN$ acceptors are generated in a cube of size $L = (N/N_D)^{1/3}$ by means of the random numbers generator. Coulomb energy of the system is calculated by randomly distributing of $(1 - K)N$ electrons over the donors. The minimal energy of the system, corresponding to the ground state, is reached by permuting the electrons over the filled and unfilled donor atoms. For a given realization of the ground state, $\mathcal{E}^2(m)$ and $\frac{\partial \mathcal{E}_z(m)}{\partial z}$ at $m$th neutral donor are calculated by summing up the electric field and its gradient created by the $N$ charged donors and $KN$ acceptors [5, 27],

$$\mathcal{E}^2(m) = \left[ -\sum_{i=1}^{N}(1 - n_i)\frac{\mathbf{R}_{im}}{R_{im}^3} + \sum_{j=1}^{KN}\frac{\mathbf{R}_{jm}}{R_{jm}^3} \right]^2 \tag{3}$$

$$\frac{\partial \mathcal{E}_z(m)}{\partial z} = \sum_{i=1}^{N}\frac{(1 - n_i)}{R_{im}^3}\left(1 - \frac{3Z_{im}^2}{R_{im}^2}\right) - \sum_{j=1}^{KN}\frac{1}{R_{jm}^3}\left(1 - \frac{3Z_{jm}^2}{R_{jm}^2}\right), \tag{4}$$

where $n_i = 1$ if a donor is neutral and $n_i = 0$ if it is ionized; $\mathbf{R}_{im} = \mathbf{R}_i - \mathbf{R}_m$ is a distance between the $m$th donor and all other impurities. The dimensionless radius-vectors $\mathbf{R}_i$, the square of the electric field $\mathcal{E}^2$ and its derivative $\frac{\partial \mathcal{E}_z}{\partial z}$ are expressed in units $N_D^{-1/3}$, $\left(\frac{e N_D^{2/3}}{\epsilon_0}\right)^2$, and $\frac{e N_D}{\epsilon_0}$, correspondingly. The directions of the Cartesian coordinate system are chosen to be along the cube axes. $\mathcal{E}^2$ and $\frac{\partial \mathcal{E}_z}{\partial z}$ are calculated according to Eqs. (3) and (4) for each realization and averaged over these realizations. In our calculations $N = 800$ donor atoms in the cube are chosen and the parameters are averaged over 30 realizations of donor and acceptor coordinates. The electrons in the impurities are distributed by correlated manner, corresponding to zero temperature, when the acceptors are charged by taking an electron from the nearest-neighboring donors. The distribution functions of $\mathcal{E}^2$ and $\frac{\partial \mathcal{E}_z}{\partial z}$ for the sample with $N_D = 4 \times 10^{14} cm^{-3}$ and $K = 0.5$ are presented correspondingly in Figs. 6 and 7.

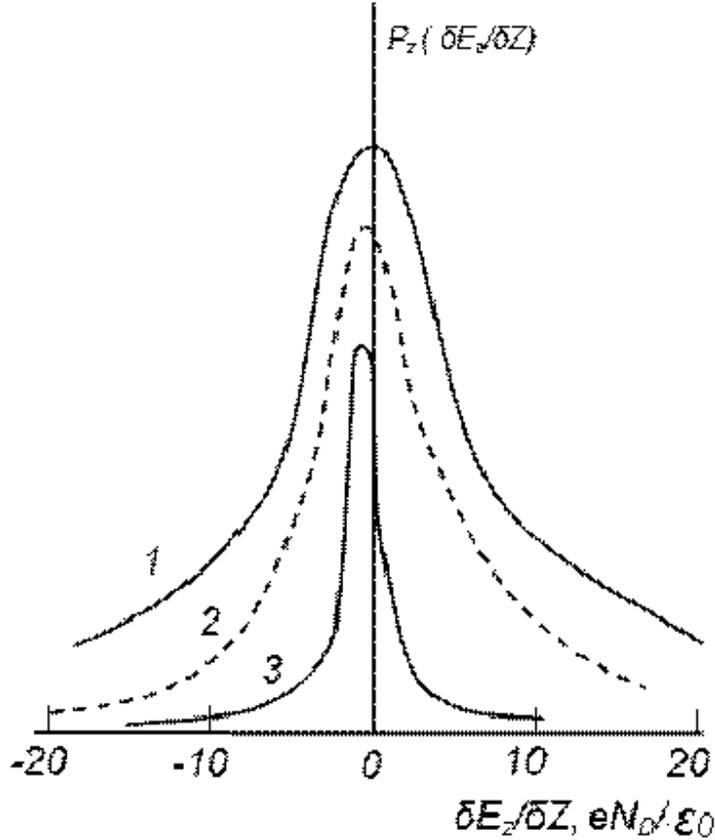

FIG. 7: Distribution of $P_2 \left( \frac{\partial \mathcal{E}_z}{\partial z} \right)$ on neutral donor at $N_D a_B^{*\,3} = 4 \times 10^{-4}$ and $K = 0.5$ for different impurity potentials (1) Coulomb potential at $T = \infty$, (2) Coulomb potential at $T = 0$, and (3) screened potential at $T = 0$ with $r_s = 10 \; a_B^*$. The gradient $\frac{\partial \mathcal{E}_z}{\partial z}$ is given in unit of $\frac{e N_D}{\epsilon_0}$.

At the breakdown electric field, free electrons with the concentration $n$, determined by the current density $j = e n \mu_c \mathcal{E}_{BD}$, are generated in the conduction band. Beside $n$ free electrons, there appear the same amount of positively charged donors, and the charge concentration becomes $N_I = 2 K N_D + n$. The free electron concentration $n$ even at strong breakdown $\mathcal{E} = \mathcal{E}_{BD}$ amounts to a few percent of the total neutral donors $n \ll N_I \approx 2 K N_D$, which was confirmed by the experimental studies of the plasma shift of the CR line in $n - GaAs$ [28].

It is necessary to take into account screening of the Coulomb potential of charged impurities by the free electrons at the breakdown. At enough high concentration $n$ of the free electrons, when the Debye radius $r_D = (\epsilon_0 k_B T / 4 \pi e^2 n)^{1/2}$ is smaller than the mean distance $R_i$ between the neutral donors and the charged impurities, $r_D \leq R_I$, the distribution probabilities $\mathcal{E}^2$ and $\frac{\partial \mathcal{E}_z}{\partial z}$ can be calculated by replacing in Eqs. (3) and (4) the Coulomb potential $e/\epsilon_0 r$ by the screened potential $\frac{e}{\epsilon_0 r} \exp\left(-\frac{r}{r_D}\right)$. This replacement yield,

$$\mathcal{E}^2(m) = \Big[ -\sum_{i=1}^{N}(1-n_i)\frac{\mathbf{R}_{im}}{R_{im}^2}\left(\frac{1}{R_{im}} + \frac{1}{r_D}\right)e^{-\frac{R_{im}}{r_D}} +$$
$$\sum_{j=1}^{KN}\frac{\mathbf{R}_{jm}}{R_{jm}^2}\left(\frac{1}{R_{im}} + \frac{1}{r_D}\right)e^{-\frac{R_{im}}{r_D}}\Big]^2 \tag{5}$$

$$\frac{\partial \mathcal{E}_z(m)}{\partial z} = \sum_{i=1}^{N} \frac{(1-n_i)}{R_{im}^3} \left( 2 - \frac{5Z_{im}^2}{R_{im}^2} - \frac{Z_{im}^2}{r_D R_{im}} - \frac{Z_{im}^2}{r_D^2} \right) e^{-\frac{R_{im}}{r_D}} -$$

$$\sum_{j=1}^{KN} \frac{1}{R_{jm}^3} \left( 2 - \frac{5Z_{jm}^2}{R_{jm}^2} \frac{Z_{jm}^2}{r_D R_{jm}} - \frac{Z_{jm}^2}{r_D^2} \right) e^{-\frac{R_{jm}}{r_D}}, \tag{6}$$

The distribution functions calculated for $n - GaAs$ at $N_D = 4 \times 10^{14} cm^{-3}$ and $K = 0.5$ are depicted in Figs. 6 and 7 for different values of the screening radius.

The line width caused by the interaction of the quadrupole moment of a neutral impurity with the electric field gradient can be calculated according to the expression [4]

$$\Delta = 2.53(\Delta Q) Ry^* (N_I a_B^{*3}), \tag{7}$$

where $\Delta Q$ is a difference of the quadrupole moments of the ground state and an excited state. Note that, a dependence of quadrupole moments of several states on the magnetic field for an isolated hydrogen-like atom is presented in Ref. [4]. Estimation of $\Delta$ for a sample with $N_D \approx 4 \times 10^{14} \ cm^{-3}$ and $K = 0.5$ in the magnetic field when $\gamma \sim 0.5$ yields $1 \ \mu meV$ and $0.7 \ \mu meV$ correspondingly for non-correlated and correlated electron distributions.

In Fig. 8 we show the dependence of the half-width $\Delta E$, caused by the quadratic Stark effect, on the screening radius, calculated for the $1s \rightarrow 2p_{+1}$ line according to the formula,

$$\Delta E = 75.75(2Ry^*)(N_D a_B^{*~3})^{4/3} \mathcal{E}^2. \tag{8}$$

All these calculations show that the distribution probabilities $\mathcal{E}^2$ and $\frac{\partial \mathcal{E}_z}{\partial z}$ are squeezed and their long tails disappear gradually with decreasing of the screening radius or increasing of the free electron concentration. In order to narrow the $1s \rightarrow 2p_{+1}$ line, e.g. 4 times the Debye radius has to be $r_D \sim 10a_B^*$ (see, Fig. 8) which corresponds to $n \sim 2 \times 10^{13} cm^{-3}$ for the free electron concentration or 10% of the total neutral donor concentration. Possible ionization of a such high concentration of neutral donors at $\mathcal{E} = \mathcal{E}_{BD}$ seems to be doubtful. Nevertheless, we have to note that (i) although the $1s \rightarrow 2p_{+1}$ line-shape was calculated for zero magnetic field, the experiments were done at strong magnetic fields when $\gamma = \frac{\hbar \omega_c}{2Ry^*} \sim 1$, and (ii) the most line narrowing occurs in the "candle"-like region of the sample CVC. It is known that a current filament with high carrier concentration, much more than the average bulk concentration, is formed in this region. We think that the main contribution to the current comes from the filaments, which determines the PES line shape at the breakdown.

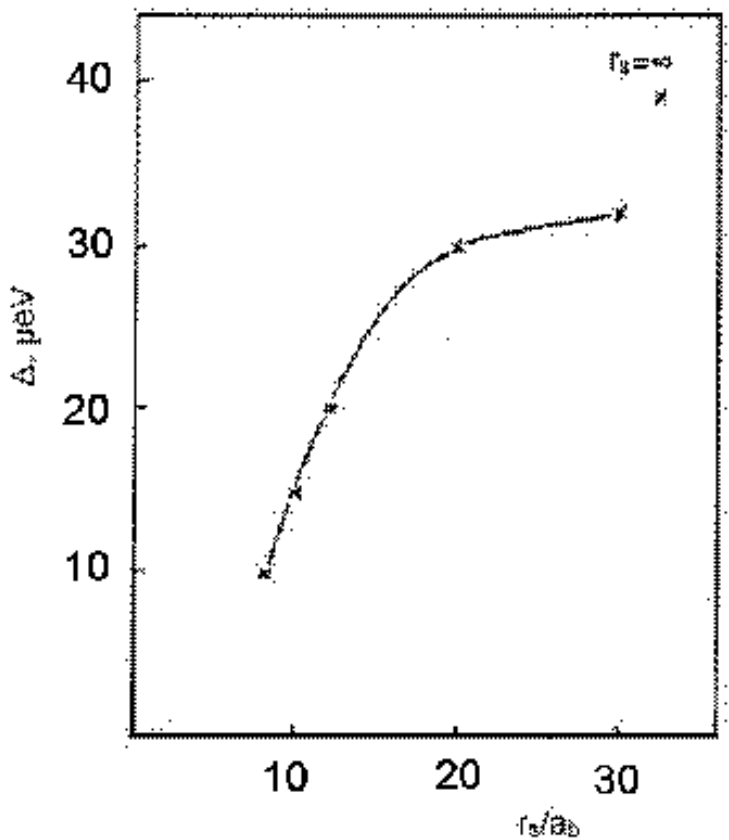

FIG. 8: The numerical dependence of the $1s \rightarrow 2p_{+1}$ line width at half-height, caused by the square Stark effect, on the screening radius at $H = 0$, $N_D a_B^{*3} = 4 \times 10^{-4}$, $K = 0.5$ and $T = 0$.

At higher concentration of the free electrons, not only the charged ions surrounded neutral impurities are screened, but the screening affects on the bound states spectrum of neutral atoms too. The screened Coulomb potential $(e/\epsilon_0 r) \exp(-r/r_s)$, in difference from the 'bare' Coulomb $e/(\epsilon_0 r)$, has the finite number of the discrete states. Investigation of the screened Coulomb potential spectrum [29] in the absence of a magnetic field yields that all discrete states higher than $N \geq 4$ disappear for the screening radius $r_s = 10 a_B^*$, and the states $N \geq 3$ disappear for $r_s = 9 a_B^*$. The quasi-discrete states of the Coulomb potential in a magnetic field are higher excited states with large radii; so they are more sensitive to the screening of free electrons. Therefore, a disappearance of the line corresponding to the transitions from $1s$ ground state to the quasi-discrete $\{N, M, \nu\}$ excited states at the breakdown electric field (see, Fig. 5) must be explained as a result of the screening.

The narrowing of $1s \rightarrow 2p_{+1}$ PES line at the breakdown electric field reveals a fine-structure of the line due to the spin-splitting of particular donor's states especially for donor with higher concentrations. Similar splitting of the shallow impurities $1s \rightarrow 2p_{+1}$ lines in $GaAs$ was observed in [30, 31]. The splitting in the samples with the parameters similar to those in our samples was reported [31] at $119\mu m$ wavelength due to the narrowing the $1s \rightarrow 2p_{+1}$ line by means of an additional illumination of the sample with a fundamental absorption edge. The authors of Ref. [30] argue that the spin-splitting of $1s$ and $2p_{+1}$ states differ each other in magnetic field with a small difference in the $1s \rightarrow 2p_{+1}$ transition energies, e.g. different $g$-factor for $1s$ and $2p_{+1}$ states, whereas a reason of the splitting is assumed in Ref. [31] to be an exchange interaction between neutral donors. We have observed this splitting in one sample at different values of the magnetic field, $H = 36.5 \ kOe$ and $H = 61 \ kOe$, by narrowing the $1s \rightarrow 2p_{+1}$ line at the breakdown electric field. The measurements at different magnetic field allow us to determine how the splitting depends on the magnetic field. The value of the splitting at $H = 61 \ kOe$ and at the absorption energy of $\epsilon_{qu} = 14.7 \ MeV$ was measured to be $\sim 25 \ \mu eV$. The splitting value can be estimated to be proportional to the difference of the Landé $g$-factors corresponding to the $1s$ ground state $g_{1s}$ and the excited $2p_{+1}$ state $g_{2p_{+1}}$, provided that the splitting is caused by the different $g$-factors, $\Delta g = \left( |g_{1s}| - |g_{2p_{+1}}| \right) = \frac{\Delta E}{\mu_B H} \approx 0.07$. This value of $\Delta g$ is very close to the difference of the $g$-factors of 0th and 1th Landau levels, determined from the CR splitting ($\Delta g \approx 0.074$) at the magnetic field $H = 61 \ kOe$ [32]. It is necessary to note that the similar splitting of the shallow donors line $1s \rightarrow 2p_{+1}$ measured by the photoexcited spectroscopy method [33, 34] has not been observed in rather pure $GaAs$ samples up to the magnetic fields $\sim 100 \ kOe$. The reason of this fact perhaps may be very close values of the $g$-factors, corresponding to the $1s$ and $2p_{+1}$ states, while both states are placed under $N = 0$ Landau level. Nevertheless, this fact does not deny to explain the line-splitting by means of the exchange interactions. Indeed, the exchange interactions should vanish at low donor concentrations due to exponentially weakening of the interactions. On the other hand the exchange interactions are sensitive to the magnetic field. We think that a clarification of a correct mechanism of the fine structure of the PES line at the breakdown needs further experimental investigations as well as theoretical studies of a magnetic field dependence of the exchange interactions.

**IIIa. Dependence of $1s \to 2p_{+1}$ line intensity on pre-breakdown electric field**

As mentioned above, PES signal at pre-breakdown electric field must be detected in constant voltage regime. PES signal is determined by $\Delta U = AU\Delta\sigma$, where A is a circuit factor, and it is not changed when the line intensity varies with voltage $U$ in all possible acceptable interval. Therefore, a dependence of the intensity on the electric field can be learned by analyzing a dependence of $\Delta\sigma(U) \sim \Delta U/U$ on applied voltage. A mechanism of the SI PES line intensity enhancement with electric field for a $1s \to 2p_{+1}$ transition close to breakdown would shed a light on the breakdown mechanism too.

An increase in the PES line intensity with the electric field has been observed and analyzed in various works [35–37], where this effect was interpreted either as increase of the hopping conductivity from the excited states [35] or with impact ionization [36] by free carriers from the excited states $2p_{+1}$, $2p_{-1}$. In Ref. [37], an increase in PES was explained by electric field ionization (tunneling) of electrons from optically excited states of SI atom to the conduction band. In [35–37] SI final states are far from the conduction band minimum, so $E_c - E_{SI}(2p_1)$ is much higher than $k_B T$. We consider the case when optically excited states $2p_{+1}$ and $3p_{+1}$ of SI are higher than $N = 0$ conduction sub-band minimum. In $n - GaAs$ such a situation is realized at higher magnetic fields satisfying the condition $\gamma = \hbar\omega_c/2Ry^* > 0.3$, when $2p_{+1}$ level crosses the $N = 0$ Landau level. In this case localized $2p_{+1}$ state is degenerated with $E(N = 0, k_Z) = \frac{\hbar\omega_c}{2} + \frac{\hbar^2 k_z^2}{2m^*}$, where $\frac{\hbar^2 k_z^2}{2m^*}$ can be estimated as difference between $E(1s \to 2p_{+1})$ transition energy of FIR laser and the value of $\hbar\omega_c/2$ for the $N = 0$ conduction sub-band (see, Fig.3). These differences for $n - GaAs$ at FIR lasers wavelengths $\lambda \approx 119\mu m$ and $84\mu m$ are about $1 meV$ and $4 meV$, respectively. This means that electric field stimulation mechanisms must be neglected in transferring of electrons from the localized $2p_{+1}$ state into the conduction band, which would take place through more fast process of recombination by emitting of acoustic and optic phonons. The $1s \to 2p_{+1}$ transition intensity was investigated for both $\mathcal{E} \| \mathbf{H}$ and $\mathcal{E} \perp \mathbf{H}$ mutual directions of electric- and magnetic fields at $H \approx 36,5\ kOe$ and $H \approx 61\ kOe$ using fixed quantum energies of FIR lasers $\hbar\omega \approx 10,45\ meV$ and $\hbar\omega \approx 14,71\ meV$, correspondingly. The intensity dependence $\Delta\sigma(U)/\Delta\sigma(0)$ on the applied voltage at $U < U_{BD}$ for $1s \to 2p_{+1}$ transitions is shown in Fig. 9. The value of $\Delta\sigma(0)$ is obtained by extrapolating of $\Delta\sigma(U)$ from small values of $U$ to zero. In order to clarify the reason of increase in $\Delta\sigma(U)$ with $U$ one considers PES signal $\Delta\sigma(U) = e\mu_c\Delta n(U)$, where $\Delta n(U)$ is the concentration of non-equilibrium electrons in conduction band at $1s \to 2p_{+1}$ photoexcitation of SI, $\mu_c$ is $N = 0$ conduction subband mobility of electrons. It is important to note that heating of electrons takes place at electric fields mach higher than brake-down field value. The criterion for heating of the electrons is that the CR line width should be asymmetric broadened due to a transition from $(N = 0, k_z) \rightarrow (N = 1, k_z)$, with $k_z \neq 0$, which takes place, as will be shown in the next Section, at fields mach stronger than the breakdown field.

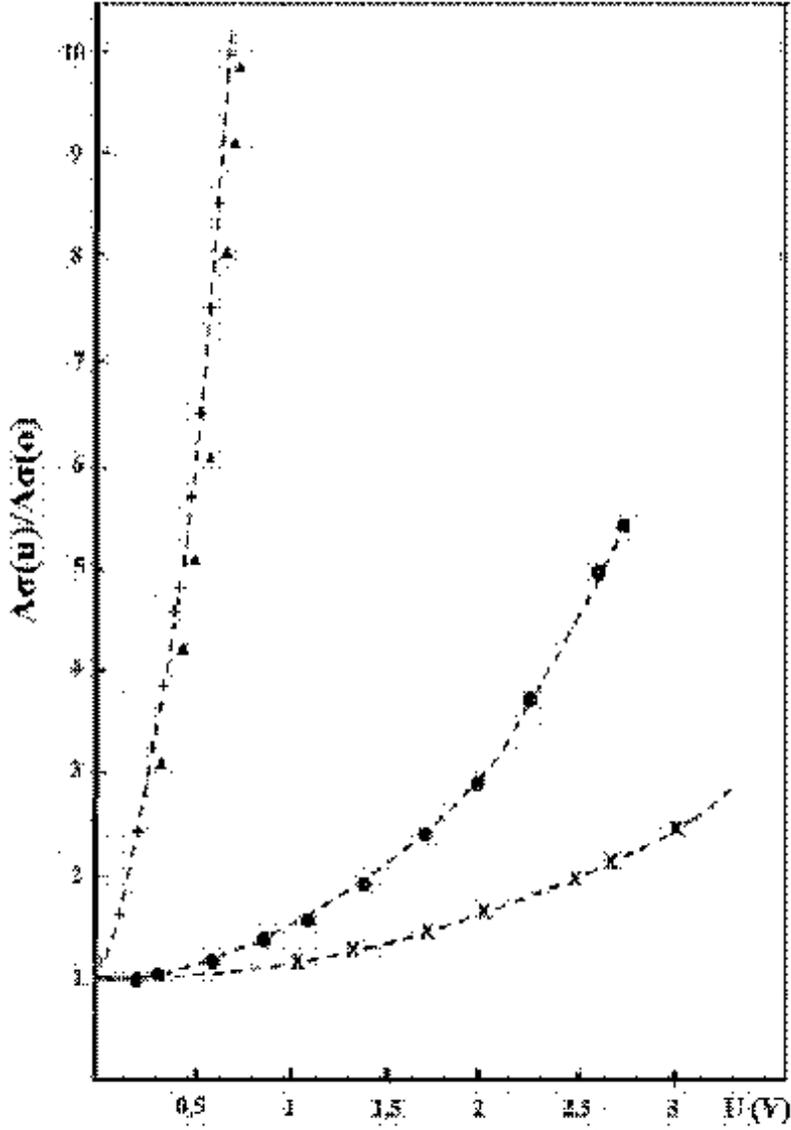

FIG. 9: Dependence of $\Delta\sigma(U)/\Delta\sigma(0)$ on the electric field $U$ at the top of $1s \to 2p_{+1}$ line at $T = 4.2\ K$ is shown for (1) $H = 36.5\ kOe$ by +-points; for (2) $H = 61.8\ kOe$ by ▲-points when $\mathcal{E}\|\mathbf{H}$ and for (3) $36,5\ kOe$ by ●-points; for (4) $H = 61,8\ kOe$ by x-points when $\mathcal{E} \perp \mathbf{H}$ in the same sample.

As it was shown in Ref. [38], heating of electrons in $n - GaAs$ samples at almost the same SI concentrations and similar to our experimental conditions takes place at electric fields much higher than those in our experiments. Concentration of nonequilibrium electrons $\Delta n(U)$ at $1s \rightarrow 2p_{+1}$ photoexcitation from SI to the conduction band can be expressed as a multiplication of the rate $G_{1s \rightarrow N=0}$ of such excitation and the time $\tau(H, U)$ necessary for an photoexcited electron to spend in conduction band, before it would be captured back

$$\Delta n(U) = G_{1s \rightarrow N=0} \tau(H, U). \tag{9}$$

Generation rate of electrons from the ground state $1s$ into the conduction band $N = 0$ through the $2p_{+1}$ state can be written as a product of the neutral donor concentration $N_D^0$, radiation intensity $P$, and the probabilities $W_{1s \rightarrow 2p_{+1}}$ and $W_{2p_{+1} \rightarrow N=0}$ for the $1s \rightarrow 2p_{+1}$ transition and the isoenergetic transition from the $2p_{+1}$ state to the $N = 0$ Landau zone, correspondingly,

$$G_{1s \rightarrow N=0} = P \, N_D^0 W_{1s \rightarrow 2p_{+1}} W_{2p_{+1} \rightarrow N=0}. \tag{10}$$

It is obvious that all above mentioned quantities do not depend on the external electric field. The only parameter which does depend under our experimental conditions on the external electric field according to [39, 40] is $\tau(H, U) = \tau_{k_z \rightarrow 0} + \tau_{cap}(H, U)$. This relaxation time consists of two parts. The first part $\tau_{k_z \rightarrow 0}$ is the relaxation time of an electron, excited from the $2p_{+1}$ localized state to the Landau subband of $N = 0$ with $k_z \neq 0$ state of the conduction subband and fell subsequently to the bottom with $k_z = 0$ of the conduction subband by many-fold irradiating acoustic phonons with characteristic energy increased in the magnetic field up to $\epsilon_{ph} = (\hbar \omega_c m^* s^2)^{1/2}$. There is no need for $\tau_{k_z \rightarrow 0}$ to depend on the external electric field because of absence of heating. The second part is the cascade capture time from the $N = 0$ subband to the SI Coulomb potential excited state settled from bottom up to $k_B T$ energy distance [39]. Then the relative increase of PES signal reads

$$\frac{\Delta \sigma(H, U)}{\Delta \sigma(H, U \rightarrow 0)} = \frac{\tau_{k_z \rightarrow 0} + \tau_{cap}(H, U)}{\tau_{k_z \rightarrow 0} + \tau_{cap}(H, 0)}. \tag{11}$$

Let us consider the relation (11) in two limiting cases, $\tau_{k_z \rightarrow 0} \gg \tau_c(H, 0)$ and $\tau_{k_z \rightarrow 0} \ll \tau_c(H, 0)$. In the first case $\Delta \sigma(\mathcal{E})/\Delta \sigma(0) \approx 1$, and there would be no electric field dependence of the

$1s \rightarrow 2p_{+1}$ line intensity. In the second case

$$\frac{\Delta\sigma(H,U)}{\Delta\sigma(H,0)} \approx \frac{\tau_{cap}(H,U)}{\tau_{cap}(H,0)}, \tag{12}$$

which means that the reason of $1s \rightarrow 2p_{+1}$ PES line intensity enhancement is an increase of the capture time in electric field. The experimental results of $1s \rightarrow 2p_{+1}$ line intensity dependence on the electric field are shown in Fig. 9. As it is seen from Fig. 9, the power-like increase in the intensity with electric field is considerably slowed down with increasing the magnetic field for a case when the electric field and magnetic field directions are orthogonal each other; whereas the intensity increase with the electric field in the case of parallel electric and magnetic fields is much sharper than that of the case of orthogonal electric and magnetic fields, and furthermore $\delta\sigma(U)/\Delta\sigma(0)$ increase with $U$ does not depend in this case on the magnetic field strength. In the first glance it seems that all results can be easily explained with carriers' heating mechanism, which is more effective in the case of $\mathcal{E}\|\mathbf{H}\|\mathbf{k_z}$ than that in the case of $\mathcal{E} \perp \mathbf{H}$. However, in order to understand the SI electric field breakdown mechanism it is important to know that the common reason of the $1s \rightarrow 2p_{+1}$ line intensity increase as well as super linear increase of the current and increase of cyclotron resonance intensity at pre-breakdown electric fields in CVC is the capture time increase (or the capture cross section decrease) with the electric field. One can conclude that, although the carriers' heating is extremely important to explain the impact ionization, this mechanism is negligible for a moderately doped $n - GaAs$ as our samples even if it takes place at all. There are two reasons, which increase the capture time in the electric field. The first one is a destruction of all discrete states of the SI Coulomb center up to the value $\epsilon_{\mathcal{E}} \approx 2(e^3\mathcal{E}/\epsilon_0)^{1/2}$ [41] and, as a consequence a reduction of the capture efficiency of the excited carriers by traps.

In difference from the $\mathcal{E} = 0$ case, electron cannot be captured when, by wandering in excited states of impurity, it appears in distance $k_BT$ below the conduction band. At $\mathcal{E} \neq 0$ it would have a probability to tunneling back by field ionization. So, the tunnel ionization in electric field will prevent diffusive descent of excited carriers to $1s$ ground state. The influence of this mechanism on capture time is more effective at highly excited states of carriers [25]. It is clear that impurity levels at which the carriers are practically captured, would be shifted down in energetical scale with increasing the electric field. We must explain why does the relative intensity decrease with increasing $H$ in the case of $\mathcal{E} \perp \mathbf{H}$, while it becomes unchanged for $\mathcal{E}\|\mathbf{H}$. Note, that this is in consistent with another experimental fact that the breakdown $\mathcal{E}$-field much weakly depends on $H$ for $\mathcal{E} \| \mathbf{H}$ as it is shown in Fig. 9.

| samples | $N_D$ $(\times 10^{14} cm^{-3})$ | $N_I = 2N_A$ $(\times 10^{14} cm^{-3})$ | $m_{CR}^*/m_0$ $\lambda = 119 \mu m$ | $m_{CR}^*/m_0$ $\lambda = 337 \mu m$ | $\Delta H_{width}$ $(kOe)$ | $\Delta H_{shift}$ $(kOe)$ |
|---|---|---|---|---|---|---|
| 1 VPE | 4.33 | 4.5 | $0.06750 \pm 0.00002$ | $0.06625 \pm 0.00002$ | 0.42 | 0.3 |
| 2 VPE | 7.35 | 9.6 | $0.06764 \pm 0.00005$ | $0.06657 \pm 0.00005$ | 0.40 | 0.28 |
| 3 LPE | 8.67 | 9.0 | $0.06777 \pm 0.00005$ | $0.06662 \pm 0.00005$ | 0.42 | 0.32 |
| 4 VPE | 1.46 | 1.2 | $0.06760 \pm 0.00005$ | $0.06635 \pm 0.00005$ | 0.42 | 0.31 |
| 5 LPE | 4.22 | 3.0 | $0.06700 \pm 0.00005$ | $0.06500 \pm 0.00005$ | 0.65 | 0.45 |
| 6 VPE | 4.31 | 4.5 | $0.06750 \pm 0.00002$ | $0.06620 \pm 0.00002$ | 0.20 | 0.18 |

TABLE II: Effective masses of $n - GaAs$ samples under investigation for the radiation wavelengths $\lambda = 119 \ \mu m$ and $\lambda = 337 \ \mu m$.

The wave functions of the SI localized states are compressed in a plane normal to the magnetic field within the cyclotron radius $l_c = (c/eH)^{1/2}$, nevertheless they are almost unchanged along the direction of $\mathbf{H}$. Therefore, it is natural to expect that depth of the SI energy states' spreading, as well as tunnel ionization in the case of $\mathcal{E} \| \mathbf{H}$ would be much more than those in the case of $\mathcal{E} \perp \mathbf{H}$. So, the relative intensity increase $\Delta \sigma(U)/\Delta \sigma(0)$ in the electric field for the $1s \to 2p_{+1}$ transition is due to decrease of capture cross section of photoexcited carriers.

# IV. CYCLOTRON RESONANCE MEASUREMENTS OF $n - GaAs$ IN DEPENDENCE ON THE ELECTRIC FIELD

## IVa. Influence of shallow impurities on the electronic effective mass in $n - GaAs$

The CR measurements were performed in $n - GaAs$ in a large interval of the electric field including the post-breakdown electric field, which allow us to understand deeper many features of the CR line-shape. The CR is well-known method to determine the carrier's effective mass $m_c^* = eH_{cr}/c\omega$ in semiconductor at the $H_{cr}$ resonance magnetic field for a given $\omega$ radiation frequency. Usually a measured effective mass differs from this definition, e. g. in $n - GaAs$, due to particularly non-parabolic zone structure and plasma oscillation of the free electrons. The cyclotron frequency $\omega_c$ is shifted from its resonance value $\omega$ due to the plasma oscillation [42] with frequency $\omega_p = (4\pi n e^2 m_0^{*-1} \epsilon_0^{-1})^{1/2}$ as,

$$\omega_c = \omega \left[ 1 - \left( \frac{\omega_p}{\omega} \right)^2 \right].$$ (13)

Then, the plasma shift of the CR line in the magnetic field at fixed value of $\omega$ is

$$\frac{\Delta H}{H_{cr}} = \frac{\Delta \omega_c}{\omega} = -\frac{\omega_p^2}{\omega^2},$$ (14)

which means that the plasma oscillation shifts the CR line to lower values of the magnetic field, which is indicated by the minus sign in the front of the last term. A correction to the cyclotron mass ($\sim \Delta H$), caused by the plasma shift, is significant at small magnetic fields and decreases as $\sim H^{-1}$ with increasing the magnetic field. Estimation of the free electron concentration from expression (14), which determine the plasma shift of the CR line, provides

$$n = \frac{m_0^* \omega^2 \epsilon_0 \Delta H}{4\pi e^2 H_{cr}}.$$ (15)

Then, the correction to the cyclotron mass can be estimated according to the above considerations to be not more than 0.03% at $H \approx 21 \ kOe$ and for $n \approx 10^{12} \ cm^{-3}$. A negligible correction to the effective mass from the plasma oscillation confirms the experimental fact that at pre-breakdown electric field no shift in the CR lines is observed while the free electron concentration increases.

Nevertheless, at the electric fields, higher than the breakdown value $\mathcal{E} > \mathcal{E}_{BD}$, and $H \approx 21 \ kOe$ a correction to the effective mass due to the plasma shift becomes essential [28]. A considerable shift in the CR line to the small magnetic field (Fig. 10) and a reduction of the cyclotron mass (Fig. 11) are observed under these conditions. Estimation of the electron concentrations according to Eq. (15) yields $n \approx 4 \times 10^{13}\ cm^{-3}$ for the measured value of shift $\Delta H \approx 0.1 \div 0.15\ kOe$ at the electric field $\mathcal{E} \approx 2\mathcal{E}_{BD}$. This value constitutes 10% of the total number of the neutral donors, under which condition the plasma corrections become observable. Note that the estimated value of the electron concentration corresponds to the current filament, therefore it may be overestimated because the CR signal is received only from the filament. A considerable asymmetric line shape of the CR to smaller values of the magnetic fields in Fig. 10 at post-breakdown fields is a result of the free electron distribution inside the filament.

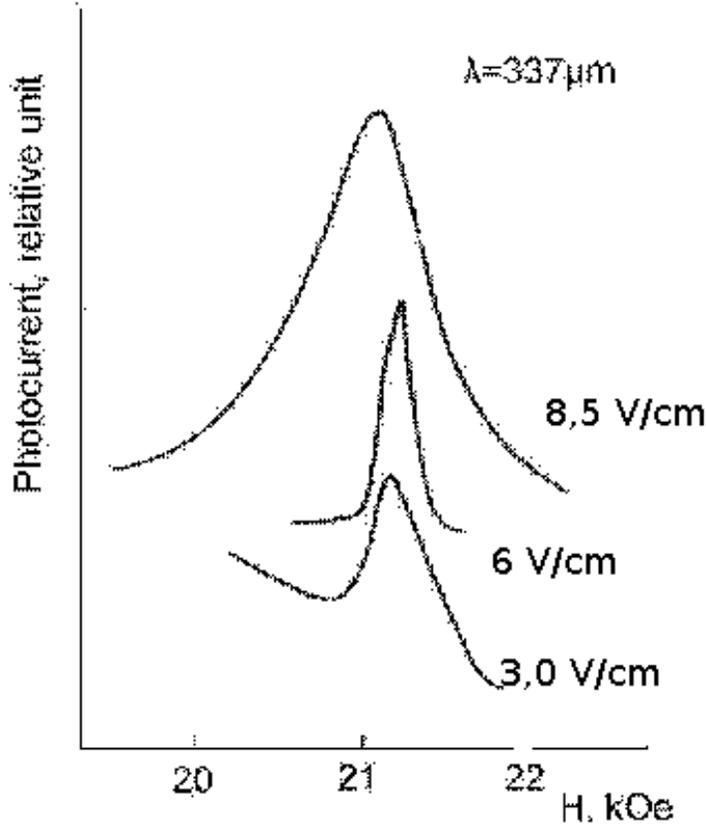

FIG. 10: Shape of the CR line for the sample I at $T = 4.2\ K$ with $\lambda \approx 337\ \mu m$ and $\mathcal{E}_{BD} \approx 5V \cdot cm^{-1}$ at different electric fields: (a) $\mathcal{E} < \mathcal{E}_{BD}$, (b) $\mathcal{E} \approx \mathcal{E}_{BD}$, and (c) $\mathcal{E} > \mathcal{E}_{BD}$. As it is seen, the plasma shift is essential for the electric field higher than the breakdown value (the case (c)), where the line shifts to the left side.

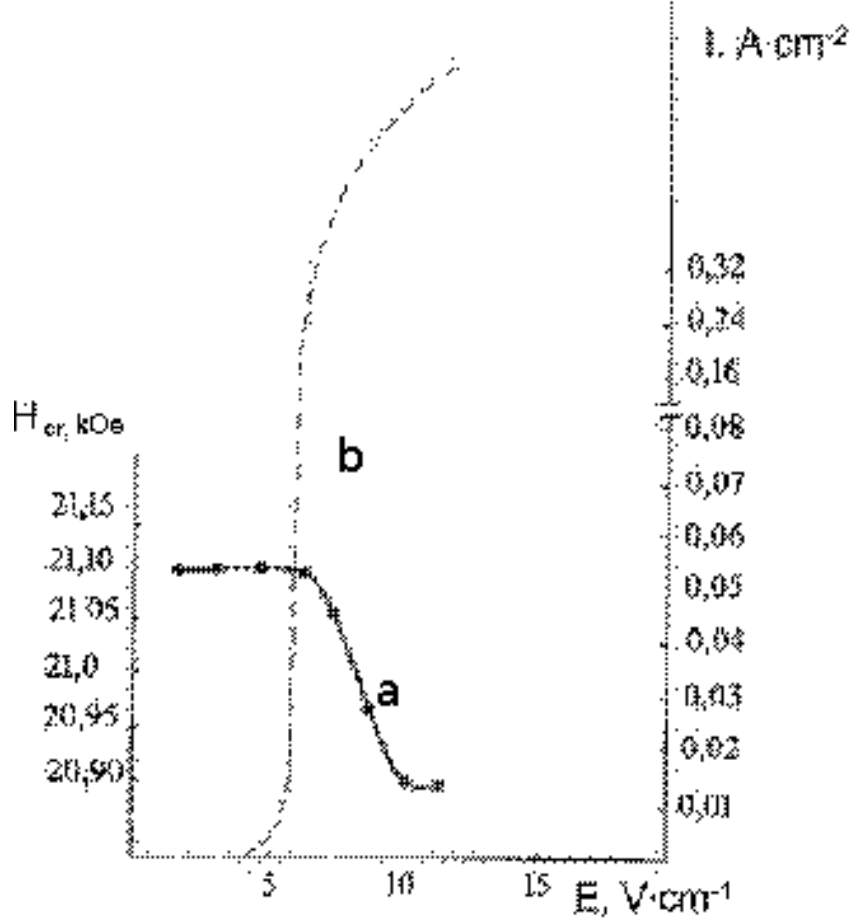

FIG. 11: (a) Plasma shift of the resonance magnetic field $H_{cr}$ in dependence on the electric field at $T = 4.2$ $K$ with $\lambda \approx 337$ $\mu m$. The curve (b) shows the CVC of the sample in the steady current regime ($R_S \ll R_L$).

Correction to the cyclotron mass $m_c^*$ due to a non-parabolic conduction band results in increase of the effective mass with increasing of the CR magnetic field [40] according to

$$m_c^*(H) = m_c^*(0) \left(1 + \frac{4\hbar\omega_c}{E_g}\right)^{1/2}.\tag{16}$$

As it is seen from Table II, the CR mass $m_c^*(61\ kOe)$ is greater than $m_c^*(21\ kOe)$ (corresponding to the radiation wavelengths $\lambda = 119\mu m$ and $337\mu m$) more than $\sim 2\%$. Nevertheless, according to Eq. (16) $m_c^*(61\ kOe) = m_c^*(21\ kOe) \cdot 1.00875$. So, assuming that the exper-imentally observed difference of the effective masses is due to the non-paraboliticity, then the difference would be 0.874 %, which is much smaller than the experimental observation. So, neither the plasma shift nor the band non-paraboliticity can not be a reason of the CR masses as in the same as well in the different magnetic fields.

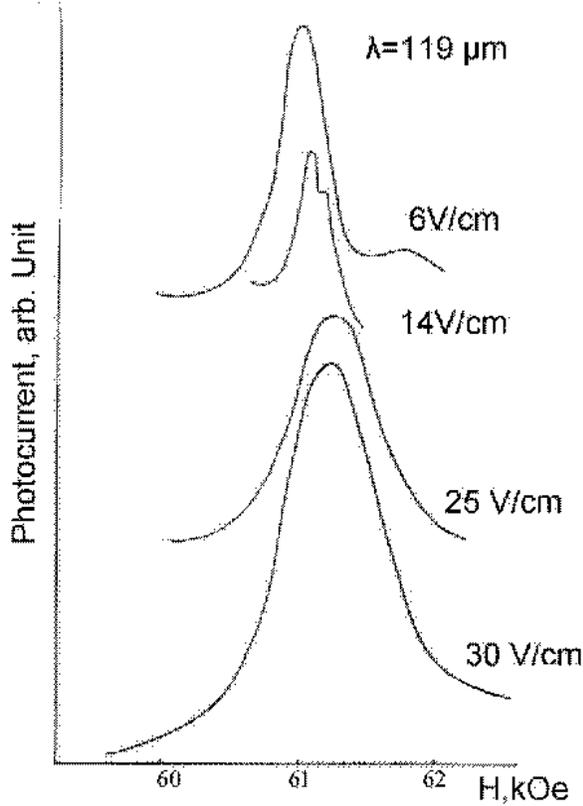

FIG. 12: Shape of the cyclotron resonance lines of the sample I at $T = 4.2\ K$, $\lambda \approx 119\mu m$, $\mathcal{E}_{BD} \approx 14\ V \cdot cm^{-1}$, and at different electric fields. The measurements were done in the voltage regime far from the breakdown, and in the current regime at the breakdown.

The effective mass in the CR experiments in $n - GaAs$ was measured by other authors too, where the measured effective masses varies in a wide interval from $m^* = 0.0665\ m_0$ [43, 44] up to $m^* = 0.07\ m_0$ [45] under the identical conditions of the CR experiments. This difference amounts 5% of the effective electron mass.

It is worthy to note that the effective cyclotron mass of electrons differs each other more that 0.5% at pre-breakdown electric field even in the samples with close physical properties, which is much more than the experimental error and it increases with the impurity concentrations.

It is necessary to stress out an existence a contribution to the effective mass due to the polaronic effects. Indeed, an electron mass is renormalized by means of the polaron constant $\alpha$ as [46, 47],

$$m_{pol} \approx m^* \left(1 + \frac{\alpha}{6}\right),\tag{17}$$

where $\alpha = \frac{e^2}{\hbar}\left(\frac{1}{\epsilon_\infty} - \frac{1}{\epsilon_\infty}\right)\left(\frac{m^*}{2\hbar\omega_{LO}}\right)^{1/2}$ with $\hbar\omega_{LO}$ being a LO phonon energy, $\hbar\omega_{LO} \approx 36~meV$. The effective dielectric constant $\epsilon^*$, existing in the polaron theory, is defined as

$$\frac{1}{\epsilon^*} = \frac{1}{\epsilon_\infty} - \frac{1}{\epsilon_0},\tag{18}$$

where, $\epsilon_\infty$ and $\epsilon_0$ are high-frequency- and static-dielectric constants, respectively. According to the percolation theory of Efros and Shklovskii [48], the static dielectric constant in a system with randomly distributed metallic and dielectric regions diverges with approaching to a critical percolation threshold value of the metallic region's volume fraction, where a insulator-metal phase transition takes place. In the insulator phase $\epsilon_0$ takes small values, comparable with $\epsilon_\infty$, yielding relatively high value for the effective dielectric constant $\epsilon^*$. In this case, the polaron constant $\alpha$ takes small values, which does not change the effective mass. The static-dielectric constant $\epsilon_0$ diverges [48] in the breakdown regime, which can be considered as a insulator-metal phase transition too. As a result, the polaron constant $\alpha$ increases due to decrease in $\epsilon^*$, yielding a considerable enhancement of the effective mass. Estimation of $\alpha$ for $GaAs$ far from the breakdown regime yields $\alpha \approx 0.06$ by taking $\epsilon_0 \approx 12.5$ and $\epsilon_\infty \approx 10.9$. The effective mass of an electron takes the value $m_{pol} = m^* \left(1 + \frac{0.006}{6}\right) \approx 0.06767$, which is in consistence with our experimental results (see, Table II). The static dielectric constant increases up to value of $\epsilon_0 \approx 10000$ inside the 'candle'-like region of CVC at the breakdown, reducing the effective dielectric constant $\epsilon^* \approx 10.9$. Therefore, the polaron mass increases as $m_{pol} = m^* \left(1 + \frac{0.32247}{6}\right) \approx 0.0706$. Such an enhancement of the effective mass around the breakdown regime may explain our experiment (see, Fig. 13), nevertheless the polaronic mechanism fails to explain a narrowing of the line-width according to Fig. 13. On the other hand, the CR line shifts to the left-side at $\lambda \approx 337~\mu m$, showing that the effective mass decreases in this case Fig. 10. This fact can not be explained too within the polaronic model. It is worthy to note that the polaronic mechanism may explain many experiments in semiconductors with ionic conductivity and in the presence of a metal-insulator phase transitions, where the dielectric constant diverges at the structural and metal-dielectric phase transitions, respectively.

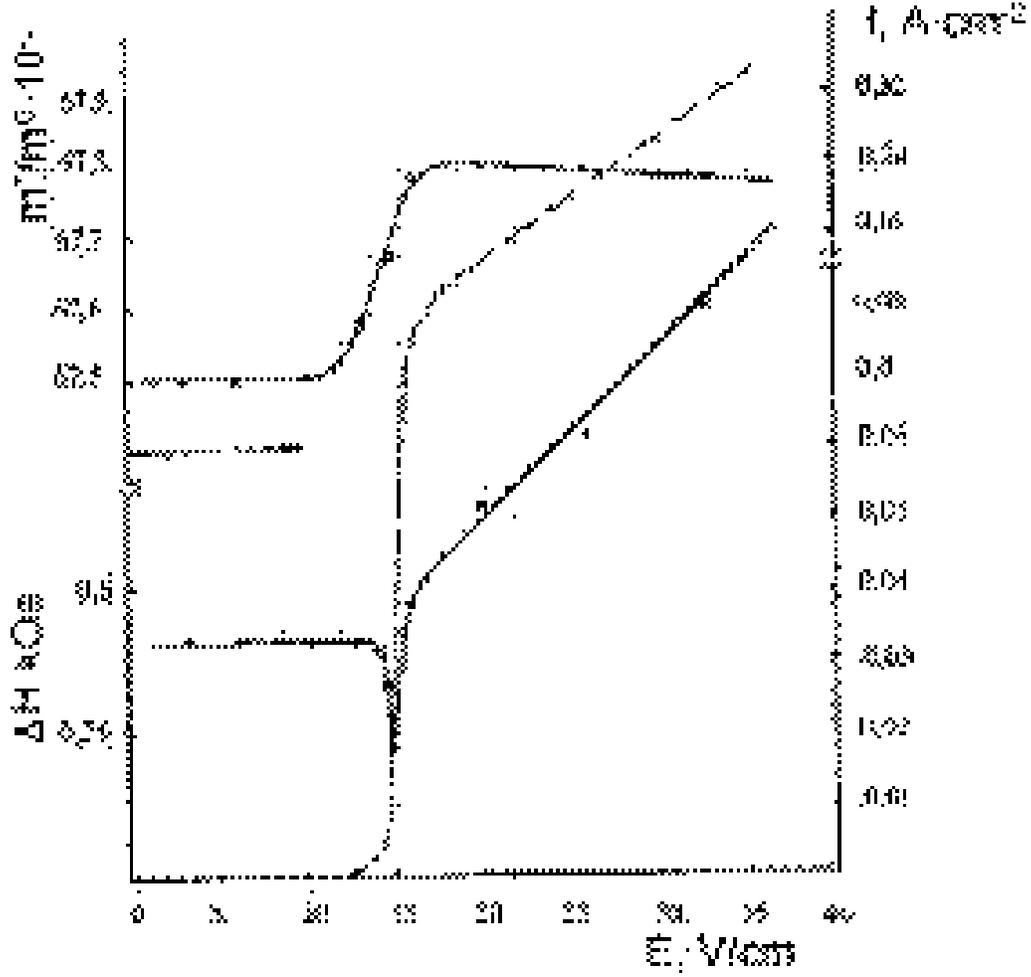

FIG. 13: Dependence of the cyclotron mass and the CR line width of the sample I on the electric field at $T = 4.2\ K$ and $\lambda \approx 119\ \mu m$. The solid dots and the open dots indicate the cyclotron mass and the line width, correspondingly. The dashed curve signifies the CVC of the sample I, which is measured in he constant current regime.

Observation of different values of the effective mass can be explained by influence of the non-homogeneous spatial distribution of the impurities on the parameters of electron in the conduction band bottom. The CR measurements in the electric field, leading to shallow impurities breakdown, confirm this assumption.

Shape of the CR line at the radiation wavelength $\lambda \approx 119\ \mu m$ for different values of the electric field, including the breakdown field, is depicted in Fig. 12. It is clearly seen from this picture that the CR lines shift to higher values of the magnetic field ($\Delta H > 0$) at $\mathcal{E} \approx \mathcal{E}_{BD}$. Although similar shift observed in e.g. $n - InSb$ [49], $CdHgTe$ [50] and $n - GaAs$ [38] crystals, has been explained as a result of electrons heating in the electric field when a resonance absorption in the non-parabolic band occurs at higher magnetic field, our investigations of the CR lines and CVC at different magnetic fields such as 21 $kOe$ ($\lambda \approx$ 337 $\mu m$), 43 $kOe$ ($\lambda \approx 172$ $\mu m$) and 61 $kOe$ ($\lambda \approx 119$ $\mu m$) contradict to the electrons heating mechanism for the CR line shift [51].

The dependence of the effective mass $m^*$ on the electric field $\mathcal{E}$ is depicted in Fig. 13. The CVC for this sample is shown by dashed curve at the magnetic field $H \approx 61kOe$, corresponding to the cyclotron resonance. A shift in the cyclotron resonance line and the corresponding increase in the effective mass is seen from the picture to take place around the breakdown electric field and it disappears at $\mathcal{E} > \mathcal{E}_{BD}$. Hence, this fact indicates that a shift in the CR line width is caused not by the electric field but by strongly increase in the free electron concentration due to breakdown of the neutral impurities at $\mathcal{E} = \mathcal{E}_{BD}$.

A shift of the cyclotron line from the resonance value $H_{cr}$ takes different values for different samples, and it amounts up to $\sim 0.6\%$ for e.g. our sample 5 at $\lambda = 119$ $\mu m$. The shift fails at relatively small magnetic fields ($H = 21$ $kOe$), and its value increases with the magnetic field from $H = 43$ $kOe$ to $H = 61$ $kOe$ [28].

The free electrons of a semiconductor, placed in the magnetic field, lie at zeroth Landau level at low temperatures, $k_B T \ll \hbar \omega_c$, and the peak in the CR (solid curve in Fig. 14) corresponds to a transition from a state with the maximal density of states (DOS) $\rho_0(E)$ in this level to the state in the first Landau level. Fluctuations of the charged impurities' density result in a randomly distributed potential of impurities, which smears the electronic states in the crystal. Therefore, the maximum of the DOS in the magnetic field shifts; the values of the shift $\epsilon_0^0$ and $\epsilon_0^1$ from the DOS maxima $\hbar \omega_c / 2$ and $3\hbar \omega_c / 2$ for a crystal depend on the impurity distribution and on the degree of non-homogeneity of this distribution. This fact allows us to determine a degree of the non-homogeneity in the impurity distribution [52]. The shift decreases with increasing the number $N$ of the Landau level due to the increase in the cyclotron radius $r_{cr}^N = [c\hbar(2N + 1)/He]^{1/2}$, i.e. the value of the DOS peak shift for zero Landau level is higher than that for the first one. Thus, the CR line peak shifts to the higher values of the energy; the value of the CR line intensity shifts for semiconductors with similar concentrations of donor and acceprots may differ each other and depend on impurity distribution in a sample. The higher the non-homogeneity of the distribution the bigger is the line shift (see, Fig. 15 and Table II). This method allows us to determine a degree of non-homogeneity in a given sample. Note that a difference between the CR masses for different samples considerably decreases at the breakdown, which supports our model.

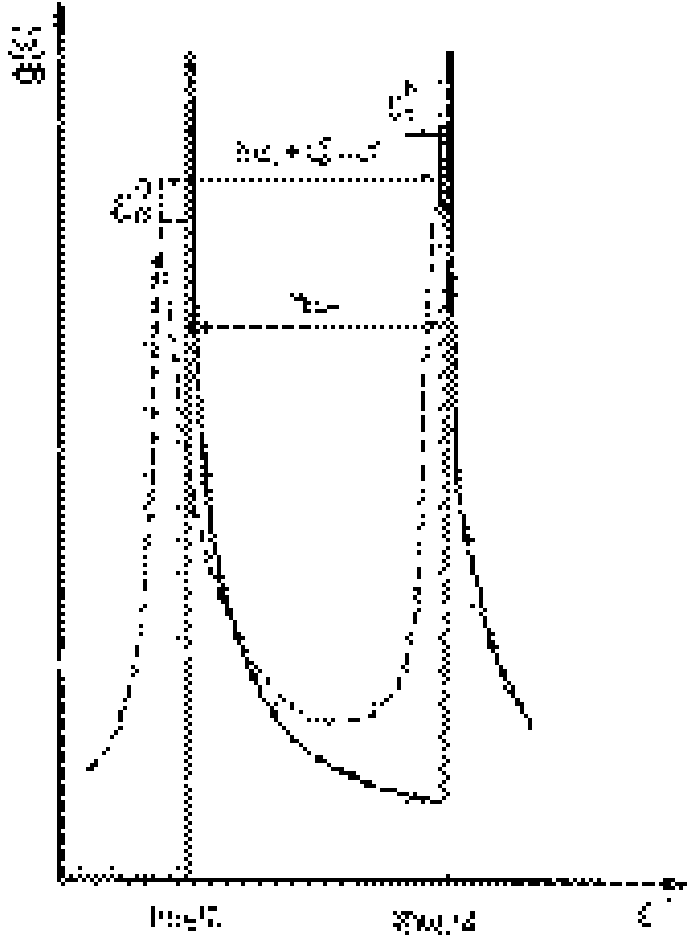

FIG. 14: The density of electronic states of the zeroth and the first Landau levels in an ideal semiconductor (solid line) and in the presence of impurities (dashed line). $\epsilon_0^0$ and $\epsilon_0^1$ denote the shifts of the zeroth and first Landau levels due to the SI fluctuation potential.

**IVb. PES Intensity Dependence of the CR Line on pre- breakdown Electric Field and influence of potential fluctuations on the CR line shape**

PES for SI $1s \rightarrow 2p_{+1}$ transition must be nonzero at arbitrary small electric field because of the final state $2p_{+1}$ in our experiments is higher than $N = 0$ Landau level. As a result, an electron photoexcited to $2p_{+1}$ appears in the $N = 0$ conduction band after its first relaxation step. However, for CR to be detected, the existence of sufficient free carrier concentration is obligatory condition. We must consider the dependence of the CR intensity on $\mathcal{E}$-field up to its pre-breakdown value $\mathcal{E} < \mathcal{E}_{BD}$. In the linear region of CVC, where the electric field induced enhancement of the free electrons is negligibly small, the free electron concentration $n(T)$ in magnetic field $H$ can be expressed ($n(T) \ll N_D$) [45] as

$$n(T) = N_c \frac{(N_D - N_A)}{2N_A} \frac{\hbar\omega_c}{k_B T} \exp\left(-\frac{E_i}{k_B T}\right), \tag{19}$$

where $N_c = 2(m^* k_B T / 2\pi\hbar^2)^{3/2}$ is the electronic density of sates in the bottom of the conduction band, $E_i = Ry^* + (\hbar\omega_c/2) - [E_{1s}(H) - E_{1s}(0)] \approx Ry^* + \hbar\omega_c/2$ is the ionization energy of SI in magnetic field which can be found from Fig. 3. The prefactor $\hbar\omega_c/k_B T$, which is added empirically to expression (19), corrects the free electron concentration in the presence of magnetic field. The amount of thermally excited free electrons at $T = 4.2$ $K$ would be only few hundred electrons by taking into account the volume of epitaxial $n$-layer of $n - GaAs$ in magnetic fields corresponding to $\gamma \approx 1$. The ionization energy of SI donors at $\gamma \approx 1$ is $E_i \approx 10.5$ $meV$ which is two times greater than $Ry^*$. This is a reason why the CR signal corresponding to the linear part of CVC is almost comparable with the noise and it is very hard to detect it not only in PES but in the transmission experiments too, without additional band gap illumination. Thus one concludes that the concentration of thermally excited free electrons in $n - GaAs$ samples at $T = 4.2$ $K$ under the magnetic field at least at $H = 61$ $kOe$ is insufficient to detect the CR. On the other hand the value of the electric field, necessary to observe the CR, increases with magnetic field approximately linearly. All these facts ensure that a necessary concentration of free electrons for observation of CR is provided by excitation of free electron from neutral impurity atoms by means of electric field. Our investigations of dependence of PES intensity of the CR spectra on the external electric field at different magnetic fields also leads to this conclusion.

Figures 16 and 17 depict a dependence of the CR intensities on the electric field for fixed radiation wavelengths $\lambda \approx 337 \mu m$ and $\lambda \approx 119 \mu m$ at magnetic fields $H \approx 21\ kOe$ and $H \approx 61\ kOe$, correspondingly, for two mutual orientations of $\mathcal{E}$ and $\mathbf{H}$ fields when $\mathcal{E} \| \mathbf{H}$ and $\mathcal{E} \perp \mathbf{H}$. It can be seen from Figs. 16 and 17 that $\Delta\sigma$ increases strongly with the applied voltage $U$ in the process of cyclotron absorption of radiation, as it takes place also in the PES of SI. Nevertheless, in difference from SI $1s \rightarrow 2p_{+1}$ PES experiments, the line intensity dependence on $\mathcal{E}$ in the CR is negligibly weak for both $\mathcal{E} \| \mathbf{H}$ and $\mathcal{E} \perp \mathbf{H}$ orientations and it does not change with increasing $H$ especially for $\mathcal{E} \perp \mathbf{H}$. There is a critical value $\mathcal{E}_{cr}$ of the electric field, behind of which the CR line intensity increases with $H$.

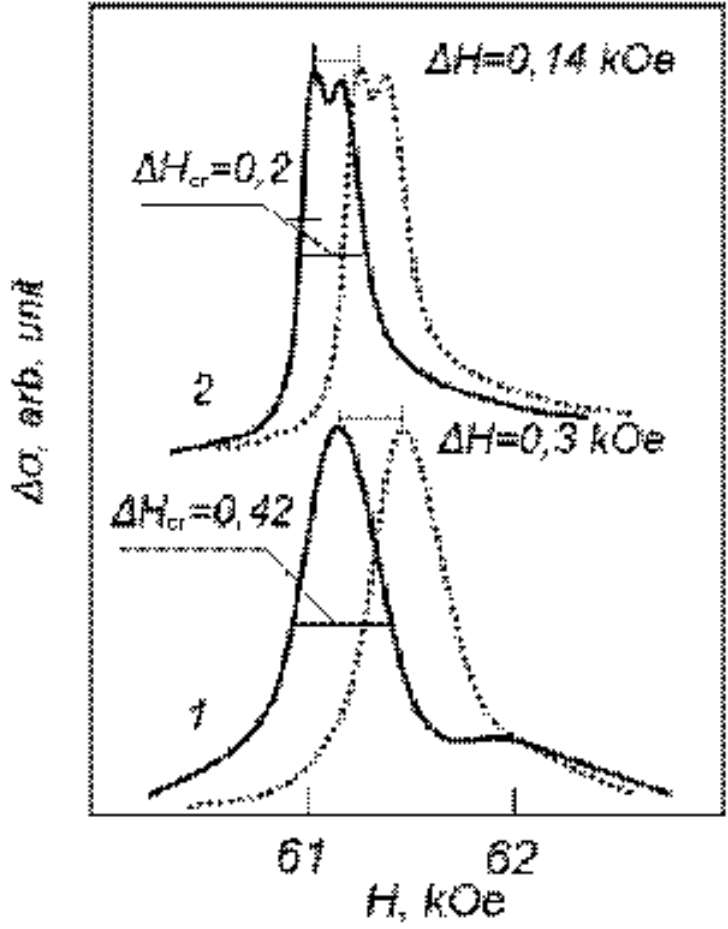

FIG. 15: The pre-breakdown line-width (solid lines) and the line-shift of the CR spectroscopy at breakdown (dashed lines) for sample 1 (curves 1) and sample 6 (curves 2) at $\lambda \approx 119 \mu m$.

In order to understand all these effects, one analyzes a change in the static conductivity $\Delta\sigma$ of samples under cyclotron absorption of the radiation. PES of the CR at pre-breakdown electric fields was investigated in constant voltage regime, so PES signal $\Delta U = AU\Delta\sigma$, where $\Delta\sigma = n_0(\mu_1 - \mu_0)J(\hbar\omega_c)$, $n_0$ is the free carrier concentration, $\Delta\mu = \mu_1 - \mu_0$ is the difference of the mobilities in the first and zeroth Landau levels (LL), correspondingly, and $J(\hbar\omega_c)$ is the radiation intensity at the CR magnetic field. The CR line intensity is determined as a product of $\Delta U/U$ and CR line width, the latter of which does not depend on electric field up to the breakdown field. In order to detect the CR in PES experiment it is necessary to determine the mobilities difference at two adjacent LL. At low temperatures, the dominant mechanism of free electron scattering is a scattering on ionized impurities ($\mu \propto \epsilon^{3/2}$). So, the mobilities' difference at CR in the absence of the external electric field ($k_z = 0$) or in the absence of free carriers' heating at pre-breakdown field is given by

$$\Delta\mu \propto \left[(3\hbar\omega_c/2)^{3/2} - (\hbar\omega_c/2)^{3/2}\right]. \tag{20}$$

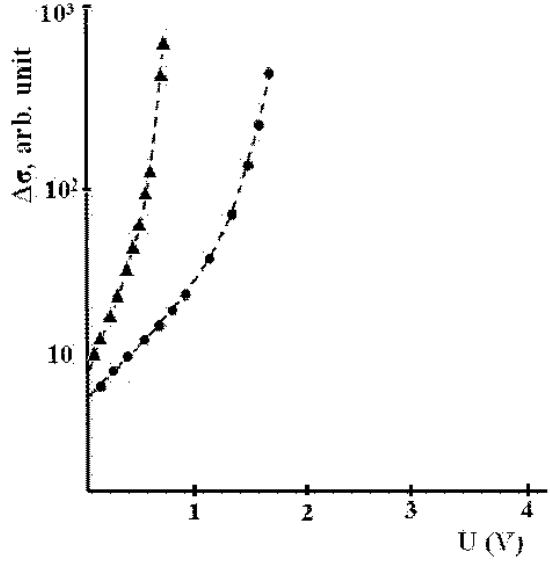

FIG. 16: Dependence of the CR PES lines of the sample 6 on the applied voltage $U$ at $T = 4.2\ K$, $\lambda \approx 337 \mu m$, $H = 21\ kOe$, where the curves marked by ▲ and ● correspond to the case of $\mathcal{E} \| \mathbf{H}$ and $\mathcal{E} \perp \mathbf{H}$, correspondingly.

This means that CR induced conductivity change is due to not free carrier concentration redistribution between LLs only, but also on mobility difference two adjacent Landau bands.

The mobility difference in electric fields leading to heating of electrons ($k_z \neq 0$) reads as

$$\Delta\mu \propto \left[(3\hbar\omega_c/2 + \hbar^2 k_z^2/2m^*)^{3/2} - (\hbar\omega_c/2 + \hbar^2 k_z^2/2m^*)^{3/2}\right], \qquad (21)$$

where the selection rules for the CR ($\Delta N = 1$, $\Delta k_z = 0$) are taken into account [53]. Supposing that the heating takes place at pre-breakdown electric fields for $H = 61\ kOe$, when the heating causes an increase in the energy $\hbar^2 k_z^2/2m^* = 30K \approx 2.5\ meV$. Then $\Delta\mu$ is calculated from the relations (20) and (21), which would differ each other only about 10%. Therefore, an increase in the line intensity is provided not by an increase in the mobility but by an increase in the free electron concentration. The above described heating of the free carriers has to increase the CR line-width asymmetrically to higher magnetic field side from the CR line maximum, which is not the case at pre-breakdown electric fields and takes place only at $\mathcal{E} > (3 \div 4)\mathcal{E}_{bd}$. This is confirmed by two facts. The first, the experimental observations, that at the breakdown electric field $\mathcal{E} = \mathcal{E}_{bd}$ a strong narrowing takes place not only in $1s \rightarrow 2p_{+1}$ line but as well in CR line-shape, refuse the heating of free carries in external electric field as a breakdown mechanism. Indeed, this fact contradicts to IIM, where heating of free carriers in pre-breakdown electric field is the necessary condition. Second, increase of CR intensity at pre-breakdown electric field is not due to increase of mobility difference $\Delta\mu = \mu_1 - \mu_0$ of free electrons at CR transition, but with increasing of carrier concentration no at zero $N = 0$ LL.

Provided that the ionization energy decreases in external electric field according to Pool-Frankel effect, the free electron concentration increase $\Delta n$ would differ from formula (19) by a factor $\beta(\mathcal{E})$ [41], which takes into account the cascade ionization of donor from $1s$ state to the excited states of SI

$$\beta(\mathcal{E}) = \frac{\beta(0)}{1 + 0.98\frac{E_{\mathcal{E}}}{k_B T}} \exp\left(\frac{E_{\mathcal{E}}}{k_B T}\right), \qquad (22)$$

where $E_{\mathcal{E}} = 2(e^3 \mathcal{E}/\epsilon_0)^{1/2}$ is an ionization energy reduction of the SI Coulomb potential in the electric field $\mathcal{E}$. Its value at breakdown electric field $E_{\mathcal{E}} \approx 0.1\ meV$ for $n - GaAs$ is much smaller than SI ionization energy $E_i$ in CR magnetic field. Two times increase of the electric field $\mathcal{E}$ from 1 to 2 $V \cdot cm^{-1}$ results in increase of CR intensity, determined by the electron concentration (19) multiplied to (22), at least 2 times while experimental increase according to Figs. 16 and 17 is much more at pre-breakdown fields. So an enormous increase of free electron concentration cannot be explained according to the ground state ionization. At low temperatures $k_B T \ll E_i$, an essential part of the carriers is localized on fluctuations of the SI Coulomb potentials $\Delta E_{fl}$, which may be identified to the half-width of the Gaussian distribution $\delta$ (see, Eqs. (1) and (2)) with the characteristic value $\Delta E_{fl} \approx 0.29 e^2 N_D^{1/3} K^{1/4} \epsilon_0^{-1}$, [54–56]. For the samples used in our experiments, with the parameters $N_D$ and $K$ given in Table I the fluctuation potential varies in the interval $\Delta E_{fl} \approx 0.1 \div 0.2\ meV$, which is smaller than $Ry^*$ and it increases with compensation $K$. Pre-breakdown electric field would easily increase the free carrier concentrations, by releasing the electrons hooked on fluctuations. The fluctuation potential distorts basically the DOS on the $N = 0$ Landau level because it has inherit properties of the conduction band in the absence of the magnetic field. Influence of the SI potential fluctuations on the conduction band for higher LLs ($N = 1, 2, \dots$) is negligibly small due to their larger CR radii. At low temperature $k_B T \ll Ry^*$, the fluctuation potential vanishes $\Delta E_{fl} = 0$ in the absence of compensating acceptors $K = 0$. Fluctuations of potential arise as a result of charged impurities influence on conduction band electrons: the negatively charged impurities increase and the positively charged ones decrease the energy of free carriers, concerning to conduction band gap at $k = 0$. In a magnetic field $H$ directed along $z$-axis CR radius for $N$-th LL is $r_{cr}^N = [c\hbar(2N + 1)/eH]^{1/2}$. Estimation of the CR radius in magnetic field corresponding to $\gamma \approx 1$ yields for zeroth and first orbits respectively $r_{cr}^0 \approx a_B^*$ and $r_{cr}^1 \approx 1.41\ a_B^*$. Note that $r_{cr}$ growth with $N$ would increase the probability of finding of charged donor-acceptor pairs, which decreases the influence of fluctuations of potential. The influence of the potential fluctuations on CR line-shape will be reduced with decreasing of the magnetic field by the same reason. Fig. 18 displays the CR line shapes for the same sample at $H = 21\ kOe$ and $H = 61\ kOe$ with increasing the electric field from its small values, corresponding to linear part of $CVC$, up to the pre-breakdown electric field in Voight geometry ($\mathbf{k} \perp \mathbf{H}$) of measurements of the CR, when applying of $\mathcal{E} \| \mathbf{H}$ is possible.

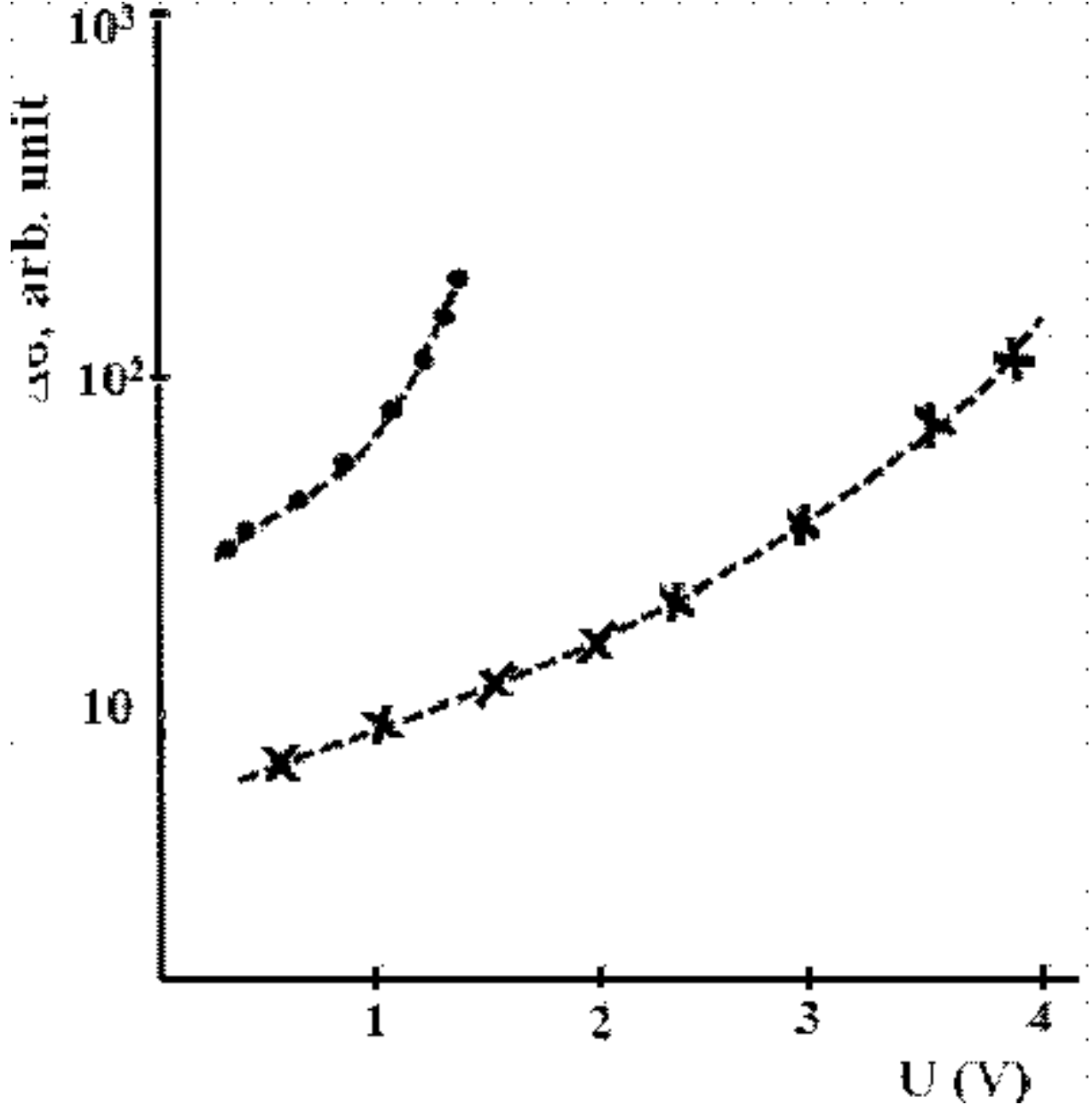

FIG. 17: Dependence of the CR PES line intensity on the applied voltage $U$ of the sample 6 at $T = 4.2\ K$, $\lambda \approx 119 \mu m$, $H = 61\ kOe$, where the curves marked by $\bullet$ and $\mathbf{x}$ correspond to the case of $\mathcal{E} \| \mathbf{H}$ and $\mathcal{E} \perp \mathbf{H}$, correspondingly.

The CR reveals a resonance absorption around the CR line from the fluctuation ptential, which has not been observed in other experiments in $n - GaAs$ samples. As it is seen from Fig. 18, apart from the CR line at $H \approx 61 kOe$ a wide absorption band is observed at higher fields, which contains several lines with low resolutions. Such kind a resonance absorption band was observed for the long wavelength illumination $\sim 337\ m\mu$ (see, Fig. 18a) of the sample 6, where less resolution results in an asymmetric broadening of the resonance CR line at higher magnetic fields. An electric field transfers the electrons from the absorption band of the fluctuation potential to the conduction band. Therefore, the intensity of the CR line increases in the electric field rather rapidly than that of the absorption band. Note that the effect in more precisely observed in the electric fields parallel to the magnetic field. The same tendency is observed also under the illumination of samples in the region of fundamental absorption in the presence of a constant electric field, Fig. 19.

Finally, we would like to discuss the low-temperature breakdown mechanism of SI by an electric field. The free carrier concentration $n$ increases in the electric field $\mathcal{E}$ not only due to an increase in the ionization coefficient $\beta(\mathcal{E})$ but also a decrease in the capture coefficient $\alpha(\mathcal{E})$. The value of $n$ can be estimated from the condition of balance between the capture and the thermal ionization as [20]

$$n(\mathcal{E})\alpha N_D^+ = \beta N_D^0,$$
(23)

where $N_D^+ = N_A + n$ and $N_D^0$ are the concentrations of charged and neutral donors. $\beta$ increases with electric field and diverges at $\mathcal{E} = \mathcal{E}_{BD}$, instead $\alpha$ decreases with increasing $\mathcal{E}$ and vanishes at $\mathcal{E} = \mathcal{E}_{BD}$, resulting in sharp enhancement of the electron concentration

$$n(\mathcal{E}) = \frac{N_D^0(\mathcal{E})}{N_D^+(\mathcal{E})} \frac{\beta(\mathcal{E})}{\alpha(\mathcal{E})}.$$
(24)

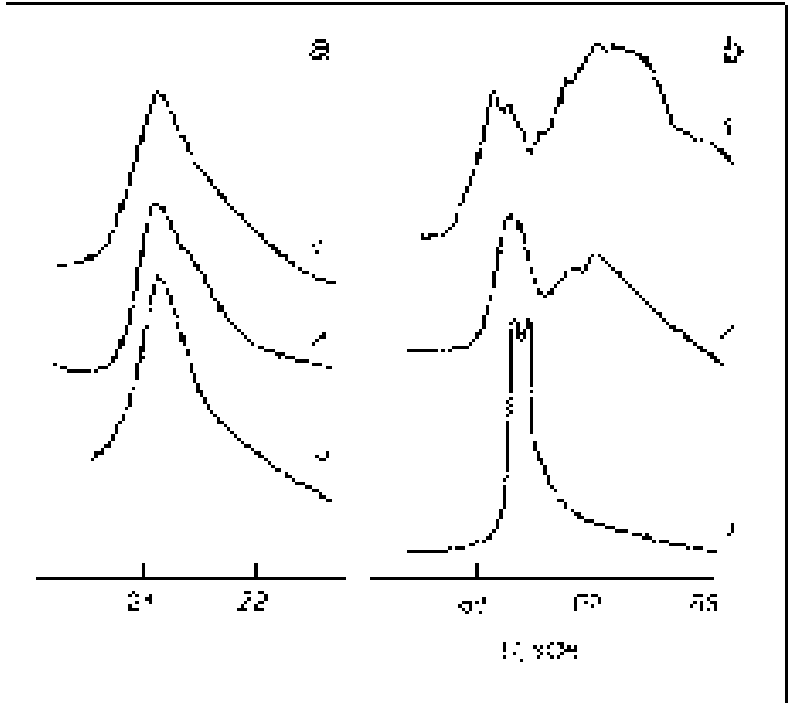

FIG. 18: Low temperature ($T = 4.2\ K$) structures of the CR lines, measured for the sample 6 at different electric fields $\mathcal{E}_1 < \mathcal{E}_2 < \mathcal{E}_3 < \mathcal{E}_{BD} = 2\ V/cm$ corresponding to the curves (1), (2), and (3), and aligned parallel to the magnetic field $\mathcal{E} \| H$, for (a) $\lambda \approx 337\ m\mu$ and (b) $\lambda \approx 119\ m\mu$.

The concentration $n$ reaches a critical value by approaching the breakdown electric field, when the charged donors become strongly screened as $\frac{e}{\epsilon_0 r} e^{-r/r_s}$. Further increase in $n$ decreases the capture cross-section of the charged donor drastically because of screening, resulting in avalanche increase of the concentration. So, the current shows an avalanche-like increase with the electric field as $j(\mathcal{E}) = en(\mathcal{E})\mu(\mathcal{E})\mathcal{E}$. We should to note that the above described breakdown mechanism is valid for the intermediate concentration $N_D = 10^{14} \div 10^{16} cm^{-3}$ of SI in a crystal. For higher concentrations of the impurities, the screening radius decreases approaching to the Bohr radius, $r_s \approx a_B^*$, when all impurities become ionized. The condition $r_s \approx a_B^*$ corresponds also to Mott transition when $a_B^* N_D^{1/3} \approx 0.25$ is satisfied for the donor concentration $N_D \approx 2 \times 10^{16} \ cm^{-3}$. The CVC characteristics of the semiconductor in this case turns from a 'candle'-like shape to a super-linear dependence on the electric field. This means that the breakdown should not be observed for samples with the donor concentrations higher than $N_D = N_D^{cr} \approx 2 \times 10^{16} \ cm^{-3}$.

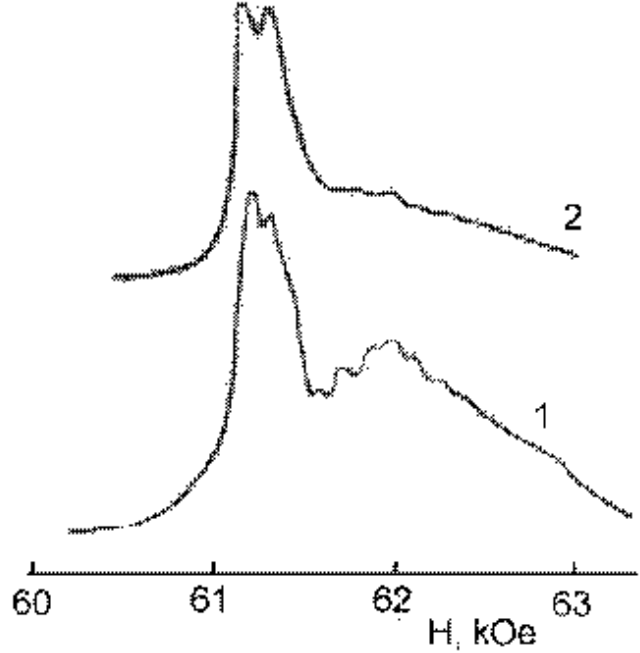

FIG. 19: Structure of the CR line of the sample 6 at $\lambda \approx 119 \ m\mu$ and $T = 4.2 \ K$. (1) $\mathcal{E} = 1 \ V/cm$ without an illumination of the sample, and (2) $\mathcal{E} = 1 \ V/cm$ under illumination in the region of fundamental absorption.

Appearance of a fluctuation band can be explained providing a non-homogeneous distribution of the impurities in a semiconductor. Fluctuations in the impurities' distribution provide a realization of a specific potential $V(\mathbf{r})$, so-called a fluctuation potential. This random potential is assumed to distribute according to Gauss law, [54, 55]

$$W(V) \propto \exp\left(-\frac{\int V^2(\mathbf{r})d\mathbf{r}}{2\gamma}\right),$$ (25)

where $\gamma$ is the mean-square fluctuation potential, which characterizes a non-homogeneous distribution of impurities. Theoretical studies of a free electron motion in the field of a random potential like 25 and under a magnetic field show a possibility of bound-electronic states below the conduction band bottom. A resonance absorption of a radiation was shown [55] to be possible in the dipole approximation by electrons bonded in the fluctuation potential bottom. Furthermore, the energy of the resonance absorption differs from that of the cyclotron absorption $\hbar\omega_c$ and this difference increases with the magnetic field.

Majority of the conduction band's electrons are localized at low temperature and in the presence of a magnetic field in the dips of the fluctuation potential. The number of electrons in the conduction band $N = 0$ increases with the electric field due to excitation of electrons from the neutral impurities, which results in enhancement of the CR intensity. On the other hand, the capture probability of the band electrons in the dips of the fluctuation potential decreases with increasing the electric field, since the latter suppresses the fluctuation dips and releases the electrons from these dips.

The similar picture is realized in the process of the illumination, which increases again the band electrons due to band-to-band excitation of electrons by illumination.

One can conclude that a non-homogeneous distribution of the impurities in the sample and its fluctuation potential results in the alternative resonance lines near the CR line.

### IVc. Effects of the electric field on the CR line width

Dependence of the CR lines on the magnetic and electric fields, temperature, the concentrations of neutral and ionized impurities yields an essential signature on the broadening mechanism of the lines. If the CR line broadening is determined by scattering of the current carriers in a semiconductor, then the momentum relaxation time $\tau_{CR}$ is expressed through the resonance magnetic field $H_p$, the half-width $\Delta H_{1/2}$ and the radiation frequency $\omega$ as [57]

$$\tau_{CR} = \frac{2H_p}{\Delta H_{1/2}\omega}.$$ (26)

Although $\tau_{CR}$ is determined by all scattering mechanisms, the main scattering mechanisms in $n - GaAs$ at $T = 4.2\ K$ are scattering on acoustic phonons as well as on neutral and ionized impurities. Since the number of the ionized impurities is much more than that of the neutral impurities, one can neglect scattering on the neutral impurities. Two experimental facts demonstrate that the scattering on the ionized impurities dominates over scattering on the acoustic phonons. First, the electron mobility $\Delta\mu = \mu_1 - \mu_0 > 0$ is changed in the cyclotron absorption, which is testified by equal polarities of the CR photo-signal and of the photoexcited impurities. The energy dependencies of the mobility for ion $\mu_{ion} \sim E^{3/2}$ [47] and acoustic phonon $\mu_{ph} \sim E^{-1/2}$ [58] scattering mechanisms differ each other. Since the electron energy increases amount of $\hbar\omega_c$ in the CR, which coincides with the ion-scattering mechanism when $\Delta\mu > 0$ and contradicts to the phonon scattering mechanism when $\Delta\mu < 0$. Second, according to the perturbation theory, the CR line should be narrowed with increasing the magnetic field either $\sim H^{-1/4}$ [59] or $\sim H^{-1}$ [60, 61] for the dominant ion-scattering mechanism, and it should broaden as $\sim H$ [58, 62] for the dominant phonon-scattering mechanism. The CR line measurements in $n - GaAs$ [24, 63] with the impurity concentration one order smaller than that in our samples were reported to show the line narrowing in the magnetic fields from $21\ kOe$ to $61\ kOe$, confirming that a scattering on the ionized impurities is a dominant mechanism at $T = 4.2\ K$ even at low impurity concentrations. Note that the charged impurities either scatter the electrons or their potential shifts the cyclotron frequency of the electrons. The relaxation time for the non-adiabatic electron scattering on the charged impurities, when the cyclotron radius $l_c$ is smaller that the radius $a$ of an impurity potential, $l_c < a$, was calculated [60] to be

$$\tau^{-1} = \frac{2\pi e^4 N_I}{\hbar\omega_c \epsilon_0^2} \left(2m^* k_B T\right)^{-1/2}. \tag{27}$$

On the other hand, a broadening of the CR line [59], due to the cyclotron frequency shift $\Delta\omega_c = (V_{xx} + V_{yy})/2m^*\omega_c$ with $V_{xx}$ and $V_{yy}$ being the second derivative of the impurity potential at the cyclotron orbit center, is given as,

$$\tau^{-1} = 0.915 \left(\frac{e^2}{\epsilon_0}\right) m^{*\ -1/4} \omega_c^{1/4} \hbar^{-3/4} N_I^{1/2}. \tag{28}$$

The relaxation time, determining from the CR line width according to Eq. (26), seems to be one order higher than that determining from Eqs. (27) and (28). One of the peculiarity of the CR measurements in $n - GaAs$ is that there is not a monotonically dependence of the line width on the ionized impurities concentration. The CR lines in Fig. 20 given for two samples with similar physical parameters show an increase in the line width with decreasing the impurity concentrations. As it is seen from these curves, the CR line of sample 5 is wider than that of sample 6, though the concentration of the ionized impurities in sample 5 is less than that in sample 6.

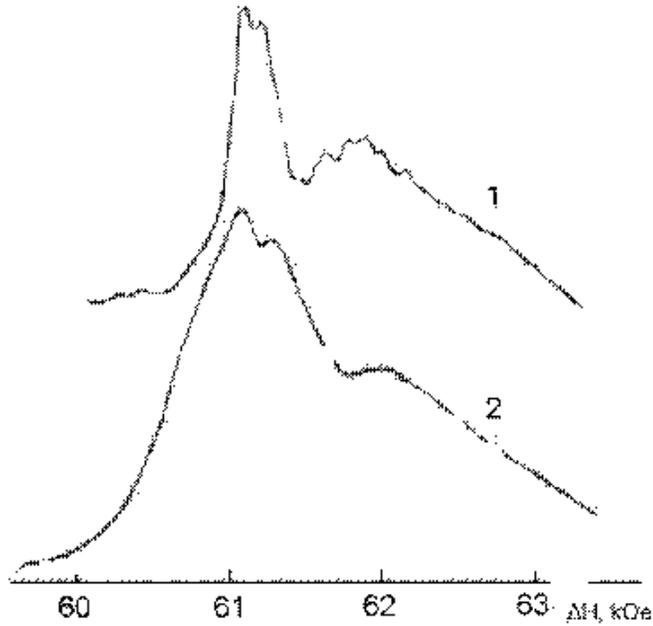

FIG. 20: CR line shapes of two samples, 5 (curve 2) and 6 (curve 1),with similar values of the neutral and ionized impurity concentrations. The electric field takes pre-breakdown values; $\lambda \approx 119\ m\mu$ and $T = 4.2\ K$. (1) $N_D = 4.3 \times 10^{14}\ cm^{-3}$ and $N_I = 4.5 \times 10^{14}\ cm^{-3}$; (2) $N_D = 4.2 \times 10^{14}\ cm^{-3}$ and $N_I = 3.0 \times 10^{14}\ cm^{-3}$.

The considerable fact is an existence of a correlation between the width and the value of shift of the CR lines at the pre-breakdown electric fields. A broadening of Landau levels due to the impurity potential fluctuations results in an additional broadening of the CR lines and shifting of their peak positions. Therefore, the CR line widths of two samples with the same concentrations of the ionized impurities can differ each other, so that the higher is the non-homogeneity of the impurity distribution the wider is the CR line.

The CR line width is defined in the energetic unit as $\Delta E = \frac{\partial E}{\partial H} \Delta H$, where the change rate of the cyclotron energy with a magnetic field $\frac{\partial E}{\partial H}$ can be estimated to be for the first two Landau levels 0.17 $meV/kOe$. This value is comparable with fluctuation potential of the charged impurities, the correct expression of which is given for a small compensation as [22],

$$E_0 = 0.29e^2 N_A^{1/3} K^{1/2} \epsilon_0^{-1}. \tag{29}$$

This expression takes a value of $0.1 \div 0.2$ $meV$ for the samples under investigation.

Broadening of the Landau levels due to the potential fluctuations results in a broadening of the CR line in addition to the broadening due to the scattering of electrons on the ionized impurities. Therefore, the wider is the CR line for the samples with equal concentration of the charged impurities the higher the non-homogeneous distribution of impurities is.

Influence of the distribution non-homogeneity on the width and the peak position of the CR line is essential only for higher magnetic fields when the cyclotron radius of the zero-th Landau level $l_c^0$ is smaller than the potential size $a_B^*$ of an impurity, $l_c^0 < a_B^*$. At smaller $H$ a cyclotron orbit with larger radius contains several charged impurities with opposite signs, the effects of which cancel each other in the averaging.

Th free electron concentration increases several order at the breakdown electric field, corresponding to a sharp enhancement of the current at $\mathcal{E} = \mathcal{E}_{BD}$ in the CVC, which changes strongly the CR line shapes. The CR line narrows approximately two times in the close vicinity of the pre-breakdown electric field (see, Figs. 12 and 13), which is observed at the radiation wave length $\lambda \approx 119\mu m$ as well as at $\lambda \approx 337\mu m$ with increasing the electric field. Recalling that the CR line broadening is caused by the ionized impurity potential, the narrowing the line width at the breakdown field can be understood as a reduction of the potential fluctuations due to the screening of the ionized atoms by free electrons released at the breakdown. The photoelectric spectrum of the CR lines, presented in Fig. 21, is measured in the steady current regime with its different values and at $\lambda \approx 337\mu m$ , which shows instead a broadening approximately three times with increasing the value of the current. The similar measurement at $\lambda \approx 119$ $\mu m$ ($H \approx 61$ $kOe$) at the breakdown field $\mathcal{E} \approx \mathcal{E}_{BD} = 15$ $V \cdot cm^{-1}$ yields the similar results. Since the broadening occurs at a steady current, it rules out the impact ionization mechanism by means of the electron heating. The theories of the CR line broadening, suggesting either $\Delta H_{1/2} \sim < \omega\tau > \sim N_I^{-1/2}$ [59] or $\Delta H_{1/2} \sim < \omega\tau > \sim N_I^{-1}$ [60, 61] on the base of the impurity ionization mechanism, can not explain the measured broadening, the value of which needs several times increase in the concentration. The observed broadening of the CR line seems to be caused by electron-electron correlations, which act similar to the scattering on the ionized impurities. Indeed, an increase in the free electron concentration with the current $J$ of cross-section area $S$ can be estimated according to

$$n \sim \frac{J}{S} \frac{1}{e\mu\mathcal{E}_{BD}}. \tag{30}$$

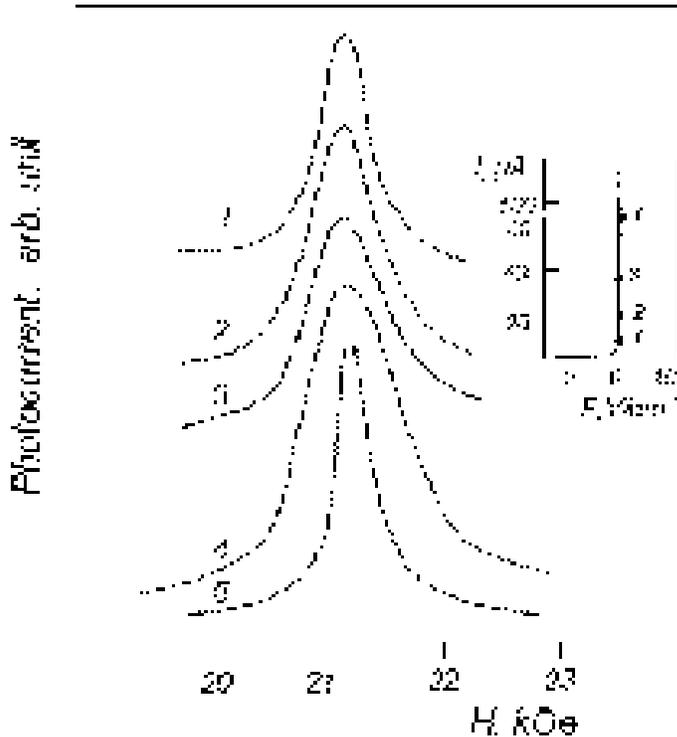

FIG. 21: CR line shape measured in the constant current regime at $\mathcal{E} \approx \mathcal{E}_{BD}$ and $\lambda \approx 337~\mu m$. The corresponding values of the current is shown in the inset of the CVC curve. The value of the current corresponding to the curve 5 is the same as in the curve 4, but the radiation intensity is 7 times lower.

The free electrons of a small concentration at the beginning step of the breakdown screen the charged impurities potential, weakening their effect on the CR line broadening. Further increase in the free electrons concentration changes the scattering mechanism from the electron-impurity to the electron-electron mechanism. The free electrons concentration in the current filament reaches the ionized impurity concentration which is several order higher than that of the average bulk concentration of the impurities.

One of the experimental evidence is that the CR line width in the broadening regime depends also on the radiation intensity. The radiation intensity for the CR line, shown by the curve 5 in Fig. 21, is 7 times higher than that of the curve 4, which results in strongly reduction in the line width. In this case, the free electron concentration considerably decreases due to an increase in the current filament cross-section resulting in a narrowing of the line width. Note that an asymmetry in the line width broadening is observed at higher breakdown fields. The asymmetry at $\mathcal{E} \approx 2\mathcal{E}_{BD}$ reaches 20% around the line maximum (Fig. 12), which seems to occur due to an electron heating by high electric field. The electron heating enhances the electron concentration with $k_z \neq 0$. Thus, the resonance condition $\hbar\nu = E(1, k_z) - E(0, k_z)$ for a non-parabolic zone is satisfied at higher magnetic fields, yielding an asymmetric broadening.

## V. CONCLUSIONS

In this paper we suggest a new mechanism of SI breakdown in $n - GaAs$ samples particularly for an intermediate donor concentration $10^{14} cm^{-3} < N_D < 10^{16} cm^{-3}$ and for the compensation degree $0.3 < K = \frac{N_A}{N_D} < 0.8$ when the electric field is varied in wide interval from the linear part of CVC up to 3-4 times higher than the SI BD field. All the obtained results contradict the impact ionization mechanism of the breakdown. The main contradiction with the IIM is the absence of carrier heating at pre-BD and BD electric fields. In difference from the $N$-like CVC dependence, which is observed in constant voltage regime [64], $S$-like CVC dependence should be observed in the heating mechanism due to strongly increase of carriers' mobility $\mu(E) \propto E^{3/2}$ in the scattering process of electrons from charged impurities. CVC characteristics of our samples have shown no $S$-like dependence in the constant current regime measurements. The other fact, which contradicts to the IIM, is that the BD electric field practically does not depend on the magnetic field in the case of parallel electric and magnetic fields $\mathcal{E} \| \mathbf{H}$ although the impurity ionization energy increases two times at the magnetic field $\gamma = \frac{\hbar\omega_c}{2Ry^*} \approx 1$.

We showed that the PES line-width of $1s \rightarrow 2p_{+1}$ transition is determined not only by the SI concentration and the compensation degree but also by the potential fluctuation, which is provided by non-homogeneous distribution of the charged impurities. The potential fluctuation disappears at the breakdown electric field $\mathcal{E} = \mathcal{E}_{\mathcal{BD}}$, when the released electrons screen the charged impurities, which may be caused also with quadratic-Stark effect and gradient-broadening mechanism of PES. Disappearance of Zeeman spectra, corresponding to the transitions from the ground state to highly excited Boyle-Howard's states up to $1s \rightarrow 3p_{+1}$ at the BD, also confirms the screening mechanism of the ionized impurities by free carriers at RD. The PES line-width is strongly narrowed at $\mathcal{E} = \mathcal{E}_{\mathcal{BD}}$.

For small concentration of impurities $N_D < 10^{12} \ cm^{-3}$ the above described mechanism is not valid, since the free electron concentration is insufficient for screening of the charged impurities. Nevertheless, the conventional impact ionization mechanism may work and the breakdown occurs at much higher (several order higher than that of for our samples) electric fields.

At higher concentrations of the SI, when $N_D > 10^{16} \ cm^{-3}$, the mean distance between the donors becomes comparable with the Bohr radius. In this case, all donors become ionized, and therefore a Mott's regime is realized. Therefore, further ionization of the impurities is not expected. The CVC in this case should not display a 'candle'-like jump, and the current will increase super-linearly with the electric field.

A breakdown of the SI in $n - GaAs$ under electric field was investigated in the work in two different configurations of the magnetic field, when the latter is parallel- and perpendicular ($\mathbf{H} \| \mathcal{E}$ and $\mathbf{H} \perp \mathcal{E}$) to the electric field. For $\mathbf{H} \perp \mathcal{E}$, the breakdown field $\mathcal{E}_{BD}$ strongly depends on the magnetic field at $H = 61 \ kOe$, corresponding to $\gamma = \frac{\hbar \omega_c}{2Ry^*} \approx 1$, as a result of two-times increases of the ionization energy of SI. Instead, CVC investigation for the case of $\mathbf{H} \| \mathcal{E}$ showed practically no dependence of $\mathcal{E}_{BD}$ on the magnetic field. Indeed, magnetic field does not change the wave function shape along the electric field when it is parallel to the electric field. A weak dependence on the higher magnetic field is a result of localization of the electronic wave in the perpendicular to the electric field plane, which reduces a transition of an electron to the excited states of the neighboring impurities.

Investigations of PES of $1s \rightarrow 2p_{+1}$ transition as well as CR line-shape at pre-breakdown electric fields have shown no correlation between the physical characteristics and the compensation $K$.

Our CR investigation of $n - GaAs$ samples yields an effective electron mass value, which varies in a wide interval. Moreover our measurements reveal that an effective mass of a sample even smaller SI concentration has higher than that of more dirty samples. Moreover, at the breakdown the effective masses of all samples become comparable each other, which means that the potential fluctuation plays an essential role. Therefore, in order to provide more correct data for an electron effective mass, it is necessary to measure the latter at the breakdown region.

**ACKNOWLEDGMENT**

This work was supported by the Science Development Foundation under the President of the Republic Azerbaijan-Grant No EIF-KETPL-2-2015-1(25)-56/01/1. The authors acknowledge A. Najafov for the technical supports.

————————

[1] K. Seeger, *Semiconductor Physics: An Introduction*, Springer-Verlag 9th Edition,2004, 537 p. (see, Section10 on the mechanism of the Impact ionization).

[2] E. I. Zavaritskaya, in *Elektricheskie i Opticheskie svoystva Poluprovodnikov*, Trudi Fiz. Inst. P. N. Lebedeva AN SSSR, Nauka (1966) p. 41 [ *Electrical and Oprical Properties of Semiconductors*, Proc. Ins. Physics Academy of Sciences USSR, Nauka (1966), p. 41].

[3] Sh. M. Kogan and T. M. Lifshitz, Phys. Status Solidi A **39**, 11 (1977).

[4] D. M. Larsen, Phys. Rev. B **8**, 535 (1973).

[5] Sh. M. Kogan and V. L. Nguyen, Sov. Phys. Solid State **15**, 44 (1981) [Fizika Tverdogo Tela **15**, 44 (1981)].

[6] O. Z. Alekperov, Phys. Status Sol. (b), **214**, 69 (1999).

[7] B. I. Shklovskii and A. L. Efros, Fiz. Tech. Poluprov. **14**, 540 (1980) [Sov. Phys. Tech. Semicond. **14**, 825 (1980)].

[8] J. Blinovsky and J. Mycielski, Phys. Rev. **36**, 266 (1964).

[9] J. Blinovsky and J. Mycielski, Phys. Rev. **40**, 1024 (1965).

[10] V. Yu. Ivanov, in *Physics of $A^{III}B^{IV}$ compounds*, p. 104 (1979).

[11] J. Yamashita, J. Phys. Soc. Jpn. **16**, 720 (1960).

[12] J. Spangler, A. Brandl, and W. Prettl, Appl. Phys. A **48**, 143 (1989).

[13] A. Dargys and J. Kundrotas, Phys. Stat. Solidi (b) **200**, 509 (1997).

[14] P. Rodin, U. Ebert, A. Minarsky, and I. Grechov, J. Appl. Phys. **102**, 034508 (2007).

[15] J. S. Townsend, Nature **62**, 340, (1900); Philos. Mag. **1**, 198 (1901).

[16] , A. F. Ioffe, Usphekhi Fiz. Nauk **8**, 141 (1928); [Sov.- Adv. Phys. **1**, 141 (1928)].

[17] K. G. Mc Kay and K. B. Mc Afee, Phys. Rev. **91**, 1079 (1953).

[18] S. H. Koenig and G. R. Gunther-Mohr, J. Phys. Chem. Solids **2**, 263 (1957).

[19] N. Sclar and E. Burstein, J. Phys. Chem. Solids **2**, 1 (1957).

[20] O. Z. Alekperov, J. Phys.: Condens. Matter **10**, 8517 (1998).

[21] O. V. Emelyanenko, T. S. Logunova, D. N. Nasledov, A. A. Telegin, and Z. I. Tchugueva,Sov. Phys. Tech. Semicond. **10**, 1280 (1976) [Fiz. Tex. Poluprov. **10**, 1280 (1976)].

[22] A. L. Efros, B. I. Shklovskii, and I. Y. Yanichev, Phys. Stat. Solidi (b) **50**, 45 (1972).

[23] W. S. Boyle and R. E. Howard, J. Phys. Chem. Solids **19**, 181 (1961).

[24] H. R. Fetterman, D. M. Larsen, G. E. Stillman, P. E. Tannenwald, and J. R. Waldman, Phys. Rev. Lett. **62**, 975 (1971).

[25] D. M. Larsen, Phys. Rev. B **24**, 1681 (1976).

[26] B. I. Shklovskii and A. L. Efros, Fiz. Tech. Poluprov. **13**, 3 (1979) [Sov. Phys. Tech. Semicond. **13**, 3 (1979)].

[27] A. L. Efros, W. L. Nguen, and B. I. Shklovskii, J. Phys.C: Solid State Phys, **12**, 1869 (1979).

[28] O. Z. Alekperov, V. G. Golubev, V. I. Ivanov-Omskii, and A. Sh. Mekhtiev, Phys. Stat. Solidi B **120**, 179 (1983).

[29] K. M. Raussel and R. F. O'Connel, Phys. Rev. A **9**, 52 (1974).

[30] R. A. Stradling, L. Eaves, R. A. Hoult, N. Muira, P. E. Simmonds, and C. C. Bradley, Proc. Int. Conf. on GaAs and related compounds, Boulder 1972 , p. 65.

[31] V. Yu. Ivanov and T. M. Lifshitz, Fiz. Tex. Poluprov. **17**, 2050 (1983)[Sov. Phys. Tech. Semicond. **17**, 2050 (1983)].

[32] H. R. Fetterman, D. M. Larsen, G. E. Stillman, P. E. Tannenwald, and J. Waldman, Phys. Rev. Lett. **21**, 975 (1971).

[33] D. M. Korn and D. M. Larsen, Solid State Commun. **13**, 807 (1973).

[34] T. S. Low, G. E. Stillman, O. J. Cho, H. Markos, and O. R. Calawa, Appl. Phys. Lett. **40**, 611 (1982).

[35] A. C. Carter, G. P. Carver, R. J. Nicholas, J. C. Portal, and R. A. Stradling, Solid State Commun. **24**, 55 (1977).

[36] E. A. Kurakova and V. I. Sidorov, Fiz. Tex. Poluprovod. **9**, 1286 (1975).

[37] S. M. Kogan and A. F. Polupanov, J. Eksp. Teor. Fiz. **81**, 2268 (1981).

[38] G. Lindeman, W. Seidenbusch, R. Lassing, J. Edlinger, and E. Cornik, Int. Conf. Wien, 1983, 473 p.

[39] V. N. Abakumov, V. I. Perel', and I. M. Yassievich, Sov. Phys. Tech. Semicond. **12**, 3 (1978) [Fizika i Texnika Poluprovod. **12**, 3 (1978)].

[40] B. Lax, J. Magn. and Magn. Mat. **11**, 6 (1979).

[41] V. N. Abakumov, L. N. Kresshuk, and I. M. Yassievich, Sov. Phys. Tech. Semicond. **12**, 264 (1978) [Fizika i Texnika Poluprovod. **12**, 264 (1978)].

[42] E. D. Palik, G. S. Picus, S. Teitler, and R. F. Wallis, Phys. Rev. **122**, 475 (1961).

[43] J. R. Waldman, H. R. Fetterman, R. E. Tannenwald, and C. C. Wolfe, Bull. Am. Phys. Soc. **16**, 1415 (1971).

[44] R. A. Stradling, *Research in quantum physics (Semiconductor physics)*, Ann. Res. of Clarendon Lab., Oxford 1972, 47 p.

[45] H. J. A. Bluyssen, J. C. Maan, T. B. Tan, and P. Wyder, Phys. Rev. B **22**, 479 (1980).

[46] Yu. A. Firsov *Polarons*, Nauka, Moscow, 1975, p. 553.

[47] A. I. Anselm, *Introduction to the semiconductors theory*, Nauka, Moscou 1978, 615 p.

[48] A. L. Efros and B. I. Shklovskii, Phys. Status Solidi (b) **76**, 475 (1976).

[49] O. Matsuda and E. Otsuka, J. Phys. Chem. Sol. **40**, 819, (1979).

[50] V. G. Golubev, V. I. Ivanov-Omskii, and T. I. Kropotov, Pis'ma v Zhurnal Tex. Fiz. **4**, 553 (1978) [Sov. Phys. JTP Lett. **4**, 553 (1978)].

[51] O. Z. Alekperov, V. G. Golubev, and V. I. Ivanov-Omskii, Fiz. Tekh. Polup. **17**, 155 (1983) [Sov. Phys. Tech. Semicond. **17**, 155 (1983)].

[52] O. Z. Alekperov, V. G. Golubev, V. I. Ivanov-Omskii, and A. Sh. Mekhtiev, *Method of the semiconductor materials analyzing by means of cyclotron resonance*, Patent No 3845907/31-25, Register. 21.11.1986, 1986.

[53] J. M. Ziman, *Principles of the Theory of Solids*, Cambridge University Press 1972.

[54] B. I. Shklovskii and A. L. Efros, *Electronic Properties of Doped Semiconductors*, Springer Series in Solid-State Sciences Vol. 45, 1984.

[55] L. B. Ioffe and A. I. Larkin, Zh. Eksp. Teor. Fiz. **81**, 1048 (1981) [Sov. Phys. JETP **81**, 1084 (1981)].

[56] *Problems of Semiconductor Physics*, Ed. V. L. Bonch-Bruevich, Moscow, 1957, 515 p.

[57] M. Fukai, H. Kawamura, K. Sekido, and I. Imaj, J. Phys. Soc. Jpn. **19**, 30 (1964).

[58] V. K. Arora and N. N. Spectror, Phys. Stat. Solidi b **94**, 701 (1979).

[59] H. Kawamura, H. Saji, M. Fukai, K. Sekido, and I. Imaj, J. Phys. Soc. Jpn. **19**, 288 (1964).

[60] A. Kawabata, J. Phys. Soc. Jpn. **23**, 999 (1964).

[61] Yu. A. Gurevich, Zh. Eks. Teor. Fiz. **66**, 667 (1974)[Sov. JETP **66**, 667 (1974)].

[62] V. K. Arora, M. A. Al-Massari, and M. Prasad, J. Phys.B **106**, 311 (1981).

[63] R. A. Stradling, K. Bajajk, J. Eaves, A. W. Levy, A. C. Carter, R. F. Krikman, J. R. Ramase, M. S. Skolnik, and R. J. Tidey, *R*esearch in Quantum Physics (Semiconductor Physics), Annual Report of the Clarendon Lab., Oxford 1974, p.28.

[64] S. M. Ryvkin, *Photoelectric Effects in Semiconductors*, Springer Series in Solid State Sciences Vol. XV, 1964, 402 p.